\documentclass[10pt,twocolumn,twoside]{IEEEtran}

\usepackage{epsf,times,amsmath,color}
\usepackage{epsfig,amsfonts,amsmath,graphicx}
\usepackage{subfigure,graphics,amssymb,amsxtra,color}
\usepackage{algorithm}
\usepackage{algorithmic}
\usepackage{flushend}

\usepackage[ansinew]{inputenc}

\usepackage{color,soul}
\usepackage{setspace}
\usepackage[hang,flushmargin,multiple]{footmisc}
\usepackage{epsfig}
\usepackage{wrapfig}
\usepackage{float}
\usepackage{amsfonts}
\usepackage{cite}
 \usepackage{acronym}
\usepackage{mymaths}
\usepackage{xcolor}

\newtheorem{lemma}{Lemma}

\newtheorem{proposition}{Proposition}
\newtheorem{corollary}{Corollary}

\renewcommand{\M}{\mathbf{\Phi}}

\renewcommand{\B}{\mathbf{B}}

\newcommand\rank{\mathrm{rank}}

\acrodef{CP}{cyclic prefix}
\acrodef{ZS}{zero pad suffix}
\acrodef{pdf}{probability density function}
\acrodef{iid}{independent and identically distributed}
\acrodef{BER}{bit error rate}
\acrodef{OFDM}{orthogonal frequency division multiplexing}
\acrodef{GSVD}{generalized singular value decomposition}
\acrodef{SVD}{singular value decomposition}
\acrodef{DMT}{discrete multitone}
\acrodef{ISI}{intersymbol interference}
\acrodef{ICI}{interchannel interference}
\acrodef{LOS}{line of sight}
\acrodef{NLOS}{non line of sight}
\acrodef{SNR}{signal-to-noise ratio}
\acrodef{PSNR}{peak signal-to-noise ratio}
\acrodef{SINR}{signal to interference plus noise ratio}
\acrodef{SIR}{signal to interference ratio}
\acrodef{MSE}{mean-squared error}
\acrodef{MIMO}{multiple-input multiple-output}
\acrodef{FFT}{fast Fourier transform}
\acrodef{IFFT}{inverse fast Fourier transform}
\acrodef{CDF}{cumulative distribution function}
\acrodef{CCDF}{complementary cumulative distribution function}
\acrodef{QAM}{quadrature amplitude modulation}
\acrodef{MMSE}{minimum mean-squared error}
\acrodef{LMMSE}{linear minimum mean-squared error}
\acrodef{SNR}{signal-to-noise ratio}
\acrodef{i.i.d.}{independent identically distributed}
\acrodef{SVD}{singular value decomposition}
\acrodef{MAP}{maximum \emph{a posteriori}}
\acrodef{MIMO}{multiple input multiple output}
\acrodef{OFDM}{orthogonal frequency division multiplexing}
\acrodef{CSI}{channel state information}
\acrodef{AWGN}{additive white Gaussian noise}
\acrodef{CDF}{cumulative distribution function}
\acrodef{KKT}{Karush-Kuhn-Tucker}
\acrodef{PDP}{power delay profile}
\acrodef{QPSK}{quadrature phase-shift keying}
\acrodef{CS}{Compressive sensing}
\acrodef{GMM}{Gaussian mixture model}
\acrodef{OMP}{orthogonal matching pursuit}
\acrodef{EM}{expectation maximization}
\acrodef{PCA}{principal component analysis}
\acrodef{LDA}{linear discriminant analysis}
\acrodef{IDA}{information discriminant analysis}
\acrodef{ML}{maximum likelihood}

\renewcommand{\B}[1]{\boldsymbol{#1}}

\renewcommand{\rank}[1]{\mathrm{rank} \left( #1 \right)}

\title{{Bounds on the Number of Measurements for Reliable Compressive Classification}}

\author{Hugo Reboredo,~\IEEEmembership{Student Member,~IEEE,}
        Francesco Renna,~\IEEEmembership{Member,~IEEE,}
	Robert Calderbank,~\IEEEmembership{Fellow,~IEEE,}
        and~Miguel R. D. Rodrigues,~\IEEEmembership{Senior Member,~IEEE}% <-this % stops a space
        \thanks{\footnotesize This paper was presented in part at the 2013 IEEE International Symposium on Information Theory {\cite{ReboredoISIT13}} and the 2013 IEEE Global Conference on Signal and Information Processing {\cite{ReboredoGlobalSIP13}}. The work of H. Reboredo was supported by Funda\c{c}\~{a}o para a Ci\^{e}ncia e Tecnologia, Portugal, through the doctoral grant SFRH/BD/81543/2011. 
The work of F. Renna was carried out in part when he was at the Department of Electronic and Electrical Engineering of University College London. The work of F. Renna was funded by the European Union's Horizon 2020 research and innovation programme under the Marie Sk{\l}odowska-Curie grant agreement No 655282. 
The work of R. Calderbank was supported in part by AFOSR under award No FA 9550-13-1-0076. 
The work of M. R. D. Rodrigues was supported by the EPSRC through the research grant EP/K033166/1. This work was also supported by the Royal Society International Exchanges Scheme IE120996. \vspace{0.2cm}}% <-this % stops a space

\thanks{\footnotesize H. Reboredo is with the Instituto de Telecomunica\c{c}\~{o}es and the Departamento
de Ci\^{e}ncia de Computadores da Faculdade de Ci\^{e}ncias da Universidade do Porto, Portugal 
({\sf email: hugoreboredo@dcc.fc.up.pt}).}% <-this % stops a space
\thanks{\footnotesize F. Renna is with the Department of Applied Mathematics and Theoretical Physics, University of Cambridge, United Kingdom 
({\sf email: fr330@cam.ac.uk}).}% <-this % stops a space
\thanks{\footnotesize R. Calderbank is with the Department of Electrical and Computer Engineering, Duke University, NC, USA
({\sf email: robert.calderbank@duke.edu}).}% <-this % stops a space
\thanks{\footnotesize M. R. D. Rodrigues is with the Department of Electronic and Electrical Engineering, University College London, United Kingdom 
({\sf email: m.rodrigues@ucl.ac.uk}).}% <-this % stops a space
}

\begin{document}

\maketitle

\begin{abstract}

This paper studies the classification of high-dimensional Gaussian signals from low-dimensional noisy, linear {measurements}. In particular, it provides upper bounds (sufficient conditions) on the number of measurements required to drive the probability of misclassification to zero in the low-noise regime, both for random measurements and designed ones. Such bounds reveal two important operational regimes that are a function of the characteristics of the source: i) when the number of classes is less than or equal to the dimension of the space spanned by signals in each class, reliable classification is possible in the low-noise regime by using a one-vs-all measurement design; ii) when the dimension of the spaces spanned by signals in each class is lower than the number of classes, reliable classification is guaranteed in the low-noise regime by using a simple random measurement design. Simulation results both with synthetic and real data show that our analysis is sharp, in the sense that it is able to gauge the number of measurements required to drive the misclassification probability to zero in the low-noise regime.
%The validity of the theoretical results presented is verified via simulation with synthetic data and with real data from a semi-supervised learning application in video motion segmentation.
\end{abstract}


\section{Introduction}

\ac{CS} is an emerging paradigm that offers the means to simultaneously sense and compress a signal without {significant loss of information~\cite{Candes06a,Candes06b,Donoho06b} (under appropriate conditions on the signal model and measurement process)}. The sensing process is based on {computing the inner product of} the signal of interest {with} a set of vectors, which are typically constituted randomly~\cite{Candes06a,Candes06b,Donoho06b}, and the recovery process is based on the resolution of an inverse problem. The result that has captured the imagination of the signal and information processing community is that it is possible to perfectly reconstruct an $n$-dimensional $s$-sparse signal (sparse in some orthonormal dictionary or frame) with overwhelming probability with only $\mathcal{O}\left(s\log\left(n/s\right)\right)$ linear random measurements~\cite{Baraniuk08,Candes06a,Donoho06b} using tractable $\ell_1$ minimization methods~\cite{Candes06b} or iterative methods, like greedy matching pursuit~\cite{Tropp10}. 

The focus of compressive sensing has been primarily on exact or near-exact signal reconstruction from a set of linear signal measurements. However, it is also natural to leverage the paradigm to perform other relevant information processing tasks, such as detection, classification and estimation of certain parameters, from the set of compressive measurements. One could in fact argue that the paradigm is a better fit to decision support tasks such as signal detection, signal classification or pattern recognition rather than signal reconstruction, since it may be easier to discriminate between signal classes than reconstruct an entire signal using only partial information about the source signal. 

This paper concentrates on the classification of signals from a set of compressive linear and noisy measurements. {In particular, we consider the case where signals associated to different classes lie on low-dimensional linear subspaces.} 
This problem is fundamental to the broad fields of signal and image processing~\cite{Ashok08,Baheti08,Baheti09}, computer vision~\cite{Mairal08,Hitomi11} and machine learning~\cite{Wright09,Chen12}, as pre-processing often relies on dimension reduction to increase the speed and reliability of classification as well as reduce the complexity and cost of data processing and computation. 

Compressive classification appears in the machine learning literature as feature extraction or supervised dimensionality reduction. 
For example, linear dimensionality reduction methods based on geometrical characterizations of the source have been developed, with \ac{LDA}~\cite{duda00} and \ac{PCA}~\cite{duda00} just depending on second order statistics. In particular, \ac{LDA}, which is one of the most well-known supervised dimensionality reduction methods~\cite{Fisher36}, addresses simultaneously the between-class scattering and the within-class scattering of the {measured} data. Linear dimensionality reduction methods based on higher-order statistics of the data have therefore also been developed~\cite{Chen12,Erdogmus04,Hild06,Kaski03,Liu12,Nenadic07,Tao09,Torkkola01}. In particular, an information-theoretic supervised dimensionality reduction inspired approach, which uses the mutual information between the data class labels and the data {measurements}~\cite{Chen12} or approximations of the mutual information via the quadratic R\'enyi entropy~\cite{Torkkola01,Torkkola03,Hild06} as a criterion to linearly reduce dimensionality, have been shown to lead to state-of-the-art classification results. More recently, learning methods for linear dimensionality reduction based on nuclear norm optimization have also been proposed~\cite{Qiu15}, {which have been shown to lead to state-of-the-art results for face clustering, face recognition and motion segmentation applications}. {Low-dimensional random linear measurements have also been used in conjunction with linear classifiers in scenarios where the number of training samples is smaller than the data dimension \cite{Durrant15}. In particular, \cite{Durrant15} derives bounds on the generalization error of a binary Fisher linear discriminant (FLD) classifier with linear random measurements.}

Compressive classification also appears in the compressive information processing literature in view of recent advances in compressive sensing~\cite{Davenport10,Wright09,Duarte06,Haupt06,Haupt07b,Wim12,Krishnamurthy10a,Krishnamurthy10b}. {Reference~\cite{Davenport10} presents algorithms for signal detection, classification, estimation and filtering from random compressive measurements.} 
References~\cite{Duarte06},~\cite{Haupt06},~\cite{Haupt07b} and~\cite{Wim12} study the performance of compressive detection and compressive classification for the case of random measurements. {References \cite{Krishnamurthy10a} and~\cite{Krishnamurthy10b} consider the problem of detection of spectral targets based on noisy incoherent projections. Reference~\cite{Wright09} notes that a small number of random measurements captures sufficient information to allow robust face recognition. The common thread in this line of research relates to the demonstration that the detection and classification problems can be solved directly in the measurement domain, without requiring the transformation of the data from the compressive to the original data domain, i.e. without requiring the reconstruction of the data.}

Other works associated with compressive classification that have arisen in the computational imaging literature, and developed under the rubric of task-specific sensing, include~\cite{Neifeld07,Ashok08,Baheti08,Baheti09,Ke10,Duarte-Carvajalino11,Duarte-Carvajalino13}. In particular, task-specific sensing, which advocates that the sensing procedure has to be matched to the task-specific nature of the sensing application, has been shown to lead to substantial gains in performance over compressive sensing in applications such as localization~\cite{Neifeld07}, target detection~\cite{Ashok08}, (face) recognition~\cite{Baheti08,Baheti09}, and reconstruction~\cite{Ke10}.

%Although various contributions in the literature have proposed efficient methods to effectively design linear projections in order to enhance the compressive classification performance, a theoretical analysis of the impact of such designs on the corresponding classification misclassification probability is provided only for some particular cases (see, e.g.~\cite{Nenadic07}). 
%
%Motivated by this observation, in this paper we address the following fundamental question: \emph{what is the minimum number of noisy, (random or designed) projection measurements needed for reliable classification?} 

The majority of the contributions in the literature to date has focused on the proposal of linear {measurement} design algorithms for two- and multiple-class classification problems (e.g. \cite{duda00,Fukunaga90,Chen12,Erdogmus04,Hild06,Kaski03,Liu12,Nenadic07,Tao09,Torkkola01,Zhang07}). Such algorithms  -- with the exception of two-class problems \cite{Fukunaga90,Zhang07} -- do not typically lead to closed-form {measurement} designs thereby not providing a clear insight about the geometry of the {measurements} and preventing us to understand how classification performance behaves as a function of the number of {measurements}. This paper attempts to fill in this gap by asking the question:

\vspace{0.2cm}

\emph{What is the number of measurements that guarantees reliable classification in compressive classification applications?}

\vspace{0.2cm}

We answer this question both for the scenario where the measurements are random and the more challenging scenario where the measurements are designed, when the distribution of the signal conditioned on the class is multivariate Gaussian with zero mean and a certain (rank-deficient) covariance matrix. {In addition, our answer to this question also leads to simple and insightful closed-form measurements designs both for two-class and multi-class classification problems.}\footnote{{Note that the problem of compressive classification of signals drawn from Gaussian distributions has been also considered in the preliminary papers \cite{ReboredoISIT13,ReboredoGlobalSIP13}, where the behavior of the misclassification probability in the low-noise regime was studied for the case of random measurements \cite{ReboredoISIT13} and for the case of designed measurements \cite{ReboredoGlobalSIP13}, but offering a closed-form characterization only for binary classifiers.}} {Analytical bounds on the number of measurements required for reliable classification are derived in this work for the asymptotic regime of low noise. On the other hand, the validity of such predictions also for positive noise levels is showcased by numerical results.}

We adopt this data model for three main reasons: {first, our classification problem corresponds to the Bayesian counterpart of low-dimensional subspace classification problems that are ubiquitous in practice;} then, this model often leads to state-of-the-art results in the compressive classification applications such as character and digit recognition as well as image classification \cite{Chen12,Torkkola01,Torkkola03}; in addition, this model -- which entails that the source distribution is a \ac{GMM} -- also relates to various well-known models in the literature including union of sub-spaces~\cite{Blumensath09,Eldar09}, wavelet trees~\cite{Blumensath09,Baraniuk10} or manifolds~\cite{Baraniuk09,Chen10}, that aim to capture additional signal structure beyond primitive sparsity in order to yield further gains. %For example, a (low-rank) \ac{GMM} can be seen as a Bayesian counterpart of the union of subspaces model. In fact, a signal drawn from a (low-rank) GMM lies in a union of subspaces, where each subspace corresponds to the image of each class conditioned covariance matrix in the model.\footnote{More generally, a signal drawn from a GMM model lies in a union of affine spaces rather than linear subspaces, where each affine space is associated with the mean and covariance of each class in the GMM model.} A low-rank GMM can also be seen as an approximation to a compact manifold. Compact manifolds can be covered by a finite collection of topological disks that can be represented by high-probability ellipsoids living on the principal hyperplanes corresponding to the different components of a low-rank GMM~\cite{Chen10}. 
{The framework based on GMM priors has also been used for the problem of signal recovery. In such case, the objective is not to determine from which Gaussian distribution the observed signal was drawn, but to reconstruct its value from compressive, noisy, linear measurements. Analytical bounds on the number of measurements needed for reliable signal reconstruction in the low-noise regime have been derived in \cite{RennaGlobalSIP13,RecJournal}.
}

The remainder of this paper is organized as follows: Section~\ref{par:PS} defines the problem, including the measurement model, source model, and performance metrics. Section~\ref{par:PeBound} presents an upper bound to the misclassification probability and its expansion at low noise that is the basis of our analysis. Sections~\ref{par:random} and \ref{par:2classes} derive upper bounds on the number of measurements sufficient for reliable classification. In Section~\ref{par:num_res} we report numerical results that validate the theoretical analysis with both synthetic data and real data from video segmentation {and face recognition} applications. {Section~\ref{par:discussion} contains a discussion on the impact of model mismatch in real data scenarios and,} 
finally, we draw conclusions in Section~\ref{par:conclusions}. The proofs of some of the results are relegated to the Appendices.

%\subsection{Notation}
The article adopts the following notation: boldface upper-case letters denote matrices $\left(\mathbf{X}\right)$, boldface lower-case letters denote column vectors $\left(\mathbf{x}\right)$ and italics denote scalars $\left(x\right)$; the context defines whether the quantities are deterministic or random.  $\mathbf{I}_{N}$ represents the $N \times N$ identity matrix, $\B{0}_{M \times N}$ represents the $M \times N$ zero matrix (the subscripts that refer to the dimensions of such matrices will be dropped when evident from the context) and $\mathrm{diag} \left(a_1,a_2,\ldots,a_N\right)$ represents an $N \times N$ diagonal matrix with diagonal elements $a_1, a_2,\ldots,a_N$. 
The operators $\left(\cdot\right)^T$, $\mathrm{rank}\left(\cdot\right)$, $\mathrm{det}\left(\cdot\right)$, $\mathrm{pdet}\left(\cdot\right)$ and $\mathrm{tr}\left(\cdot\right)$ represent the transpose operator, the rank operator, the determinant operator, the pseudo-determinant operator and the trace operator, respectively. $\mathrm{Null}\left(\cdot\right)$ and $\mathrm{Im}\left(\cdot\right)$ denote the null space and the (column) image of a matrix, respectively, and $\mathrm{dim}\left(\cdot\right)$ denotes the dimension of a linear subspace. We also use the symbol $(\cdot)^\perp$ to denote the orthogonal complement of a linear space. The multivariate Gaussian distribution with mean $\B{\mu}$ and covariance matrix $\B{\Sigma}$ is denoted by $\mathcal{N} \left(\B{\mu},\B{\Sigma}\right)$ and the symbol $\mathbb{P}[E]$ is used to denote the probability of the event $E$. $\log\left(\cdot\right)$  denotes the natural logarithm. For the sake of a compact notation, we also use the symbols $\mathcal{N}_i=\mathrm{Null}(\B{\Sigma}_i)$ and $\mathcal{R}_i=\mathrm{Im}(\B{\Sigma}_i)$, as well as $\mathcal{N}_{ij}=\mathrm{Null}(\B{\Sigma}_i + \B{\Sigma}_j) =\mathcal{N}_i \cap \mathcal{N}_j$ and  $\mathcal{R}_{ij}=\mathrm{Im}(\B{\Sigma}_i + \B{\Sigma}_j) =\mathcal{R}_i + \mathcal{R}_j$, where $+$ denotes the sum of linear subspaces. The article also uses the symbol $[x]^+=\max \{ x,0 \}$, {the floor operator $\lfloor x \rfloor$, which represents the larger integer less than or equal to $x$,} and the little $o$ notation where $g\left(x\right) = o\left(f\left(x\right)\right)$ if $\displaystyle \lim_{x \to \infty} {g\left(x\right)}/{f\left(x\right)} = 0$.

\section{Problem Statement}
\label{par:PS}

We consider the standard measurement model given by:
\begin{IEEEeqnarray}{c}
\label{s_model}
{\bf y} =  {\bf \Phi} {\bf x} + {\bf n},
\end{IEEEeqnarray}
where ${\bf y} \in \mathbb{R}^M$ represents the measurement vector, $\mathbf{x} \in \mathbb{R}^N$ represents the source vector, ${\bf \Phi} \in \mathbb{R}^{M \times N}$ represents the measurement matrix or kernel\footnote{We refer to ${\bf \Phi}$ as the measurement or sensing matrix/kernel interchangeably throughout the paper.} and $\mathbf{n} \sim \mathcal{N} \left({\bf 0},\sigma^2  {\bf I}\right) \in \mathbb{R}^M $ represents white Gaussian noise.{\footnote{The results presented in the remainder of the paper can be easily generalized to the case when the noise covariance matrix is a positive definite matrix $\mathbf{\Sigma}_{\mathbf{n}}$.}} %We consider both random measurement kernels and designed kernels that aim to improve the classification performance.

\renewcommand{\labelenumi}{A.\arabic{enumi}}

%We suppose that the source is described by a \ac{GMM}; the source signal is drawn from one of $L$ classes, the \emph{a priori} probability of class $i$ is $p_i$ and the distribution of the source conditioned on class $i$ is Gaussian with mean $\B{\mu}_i \in \mathbb{R}^N$ and (possibly rank-deficient) covariance matrix ${\bf \Sigma}_i \in  \mathbb{R}^{N \times N}$. 
{
We also consider that the source model is such that:
\begin{enumerate}
\item The source class $C \in \{1,\ldots,L\}$ is drawn with probability $p_i, i=1,\ldots,L$.
\item The source signal conditioned on the class $C=i$ is drawn from a multivariate Gaussian distribution with zero mean and (rank deficient) covariance matrix $\mathbf{\Sigma}_i \in \mathbb{R}^{N \times N}$.%\footnote{Throughout the paper, we will focus predominantly on the case of zero-mean classes, i.e., when $\boldsymbol{\mu}_i =\mathbf{0}, i=1,\ldots,L$. However, some of the results are also generalized for the case of nonzero-mean classes.}
\end{enumerate}}
We should point out that we use a low-rank modeling approach even though many natural signals (e.g. patches extracted from natural images, face images, motion segmentation features, handwritten digits images, etc.) are not always low-rank but rather ``approximately'' low-rank \cite{Chen10}. The justification for the use of such low-rank modeling approach is two-fold: first, a low-rank representation is often a very good approximation to real scenarios, particularly as the eigenvalues of the class conditioned covariances often decay rapidly; second, it is then standard practice to account for the mismatch between the low-rank and the ``approximately'' low-rank model by adding extra noise in the measurement model in (\ref{s_model}) (see~\cite{RecJournal}).\footnote{We also note that our analysis focuses on zero-mean models, since various datasets (e.g., face images and motion segmentation features) can be well represented via zero-mean classes \cite{Chen10,vidalTutorial}. However, some of the results in the paper can also be generalized to the case of nonzero-mean classes.}

It is assumed that the classifier -- which infers the true signal class from the signal measurements using a \ac{MAP} classifier -- is provided with the knowledge of the true model parameters{, i.e., the prior probabilities $p_i, i=1,\ldots,L$, the source covariance matrices $\mathbf{\Sigma}_i, i=1,\ldots,L$, the measurement matrix $\mathbf{\Phi}$ and the noise variance $\sigma^2$}.\footnote{{Although our analysis assumes that the classifier is given the true distributions, in Section~\ref{par:num_res}, we also conduct experiments with real datasets to assess scenarios where the classifier does not know the true distributions but rather approximate ones, that are learnt from training data.}} In particular, the signal class estimate produced by the classifier is given by:
\begin{equation}
\hat{C} = {\arg} {\max_{i \in \{1, \cdots, L\}} p(C=i | {\bf y})}  =   \arg \max_{i \in \{1, \cdots, L\}} p({\bf y} | C=i ) p_i,
\label{class}
\end{equation}
where $p(C=i | \mathbf{y})$ is the \textit{a posteriori} probability of class $C=i$ given the measurement vector $\mathbf{y}$  and $p(\mathbf{y} | C=i)$ represents the probability density function of the measurement vector $\mathbf{y}$ given the class $C=i$, which is zero-mean Gaussian, with covariance matrix $\mathbf{\Phi} \mathbf{\Sigma}_i \mathbf{\Phi}^T+ \mathbf{I}\sigma^2.$

Our objective is to characterize the number of measurements sufficient for reliable classification in the asymptotic limit of low noise, i.e. such that
\begin{equation}
\lim_{\sigma^2 \to 0} P_e = 0,
\label{eq:phase_transition}
\end{equation}
{
where $P_e$ is the misclassification probability of the \ac{MAP} classifier. {Note also that the asymptotic regime of low-noise plays a fundamental role in various signal and image processing scenarios, e.g., digit recognition and satellite data classification~\cite{Chen12}.}
 In particular, by using the law of total probability, we can write
\begin{IEEEeqnarray}{rCl}
P_e &=& \mathbb{P} [\hat{C} \neq C] = \sum_{i=1}^L p_i  \, \mathbb{P}[\hat{C} \neq i | C=i]\\
 & = & \sum_{i=1}^L p_i \int_{\mathbb{R}^M \smallsetminus \mathcal{D}_i}  p(\mathbf{y}|C=i)  d \mathbf{y},
\end{IEEEeqnarray}
where $\mathcal{D}_i$ is the decision region associated to class $i$, that is, the set of values $\mathbf{y}$ corresponding to the output $\hat{C}=i$. Moreover, we can express the set $\mathbb{R}^M \smallsetminus \mathcal{D}_i$ in terms of the unit step function $u(\cdot)$ and, by leveraging the definition of the MAP classifier in \eqref{class}, 
%Namely, the decision region $\mathcal{D}_i$ contains all the values of $\mathbf{y}$ such that $p_i  p(\mathbf{y}|C=i)  > p_j p(\mathbf{y}|C=j) $ for $j=1,\ldots, L$, $j\neq i$, or, equivalently, $\log   \frac{p_i  p(\mathbf{y}|C=i)}{p_j p(\mathbf{y}|C=j) } >0$ for $j=1,\ldots, L$, $j\neq i$. Therefore, the set $\mathbb{R}^M \smallsetminus \mathcal{D}_i$ can be expressed as the set of values $\mathbf{y}$ for which $\max_{\substack{j \\ j\neq  i }} \log \frac{  p_j p(\mathbf{y}|C=j)  }{p_i p(\mathbf{y}|C=i)}   >0$, and 
we can write the misclassification probability as
\begin{equation}
P_e =  \sum_{i=1}^L  p_i
 \int\displaylimits_{-\infty}^{+\infty}   p(\mathbf{y}|C=i) u \left(   \max_{\substack{j \\ j\neq  i }} \log \frac{  p_j p(\mathbf{y}|C=j)  }{p_i p(\mathbf{y}|C=i)}   \right) d \mathbf{y}.
 \label{eq:Perr}
\end{equation}
}
%
%
%
%
%
%
%
%
%\begin{equation}
%P_{e}   =  \sum_{i=1}^L  p_i
% \int\displaylimits_{-\infty}^{+\infty}   p(\mathbf{y}|C=i) u \left(   \max_{\substack{j \\ j\neq  i }} \log \frac{  p_j p(\mathbf{y}|C=j)  }{p_i p(\mathbf{y}|C=i)}   \right) d \mathbf{y}, 
%\label{eq:Perr}
%\end{equation}
%is the misclassification probability of the \ac{MAP} classifier and $u(\cdot)$ is the unit-step function. 
{The low-noise} characterization of the misclassification probability will be carried out both for random {measurements} and designed {measurements}.

Our characterization will also be based on the following additional assumptions:
{
\begin{enumerate}
\setcounter{enumi}{2}
\item The linear spaces $\mathcal{R}_i = \mathrm{Im}(\mathbf{\Sigma}_i), i=1,\ldots,L$ are of equal dimension, i.e. $\dim(\mathcal{R}_i) =r_{\B{\Sigma}}<N, i=1,\ldots,L$;\footnote{{Many of the results presented in this work naturally generalize to the case when the linear spaces spanned by signals in different classes have different dimensions.}}
\item The linear spaces $\mathcal{R}_i$ are independently drawn from a continuous \ac{pdf} over the Grassmann manifold of subspaces of dimension $r_{\B{\Sigma}}$ in $\mathbb{R}^N$, so that the null spaces $\mathcal{N}_i = \mathrm{Null}(\mathbf{\Sigma}_i)$ are also of equal dimension, i.e. $\dim(\mathcal{N}_i) = N-r_{\B{\Sigma}}$, and are also drawn independently from a continuous \ac{pdf} over the Grassmann manifold of subspaces of dimension $N-r_{\B{\Sigma}}$ in $\mathbb{R}^N$.\footnote{Note that this assumption on the linear spaces occupied by signals in different classes reflects well the behavior of many real data ensembles for various applications as face recognition, video motion segmentation, or digits classification~\cite{Chen12,vidalTutorial}.} 
\end{enumerate}}
The assumptions A.3 and A.4 impliy that with probability~1
\begin{equation}
\dim(\mathcal{R}_{ij})=   \dim(\mathcal{R}_i + \mathcal{R}_j) = \min \{ N, 2 r_{\mathbf{\Sigma}}  \},
\end{equation}
and
\begin{equation}
\dim(\mathcal{N}_{ij})=\dim(\mathcal{N}_i \cap \mathcal{N}_j) = [N - 2r_{\B{\Sigma}}]^+.
\end{equation}

Our characterization will also use the quantities:
\begin{IEEEeqnarray}{rCl}
\label{nodim}
R & =  & \dim (\mathcal{R}_i)  + \dim (\mathcal{R}_j)  - 2 \dim (\mathcal{R}_i \cap \mathcal{R}_j) \\
% & = & \dim (\mathcal{N}_i)  + \dim (\mathcal{N}_j)  - 2 \dim (\mathcal{N}_i \cap \mathcal{N}_j)\\
& = & 2 \dim(\mathcal{R}_{ij}) - \dim(\mathcal{R}_i) - \dim(\mathcal{R}_j) \\
  & = &  2 \min \{ N-r_{\mathbf{\Sigma}}, r_{\mathbf{\Sigma}} \}  ,
\end{IEEEeqnarray}
that relates to the difference between the dimension of the  sub-spaces spanned by source signals in classes $i$ or $j$ and the dimension of the intersection of such sub-spaces;
\begin{IEEEeqnarray}{rCl}
r_i & = & \mathrm{rank} (\mathbf{\Phi} \mathbf{\Sigma}_i \mathbf{\Phi}^T) \\
v_i & = & \mathrm{pdet} (\mathbf{\Phi} \mathbf{\Sigma}_i \mathbf{\Phi}^T), 
\end{IEEEeqnarray}
which measure the dimension of the sub-space spanned by the linear transformation of the signals in class $i$ and the volume occupied by those signals in $\mathbb{R}^M$, respectively, and
\begin{IEEEeqnarray}{rCL}
r_{ij} & =  & \mathrm{rank} (\mathbf{\Phi} (\mathbf{\Sigma}_i + \mathbf{\Sigma}_j) \mathbf{\Phi}^T)\\
v_{ij} & =  &  \mathrm{pdet} (\mathbf{\Phi} (\mathbf{\Sigma}_i + \mathbf{\Sigma}_j) \mathbf{\Phi}^T),
\end{IEEEeqnarray}
which measure the dimension of the direct sum of sub-spaces spanned by the linear transformation of the signals in classes $i$ or $j$ and the volume occupied by the {measured} signals from classes $i$ and $j$ in $\mathbb{R}^M$, respectively. 

%\commentFR{Is the following a section or a subsection? I will update the introduction accordingly}
\section{Misclassification Probability, Bounds and Expansions}
\label{par:PeBound}

The basis of our characterization of an upper bound to the number of random or designed measurements sufficient for reliable classification is an asymptotic expansion of an upper bound to the misclassification probability of the MAP classifier in (\ref{eq:Perr}). {We work with an upper bound to the misclassification probability \emph{in lieu} of the true misclassification probability, in view of the lack of closed-form expressions for the misclassification probability of the \ac{MAP} classifier.}

{In particular, the Bhattacharyya bound~\cite{Wim12} represents an upper bound to the misclassification probability associated to the binary \ac{MAP} classifier which is based on the inequality $\min\left\{a,b\right\} \leq \sqrt{a b}$, for $a,b>0$.} Then, 
it is possible to establish, by using the union-bound in conjunction with the Bhattacharyya bound, that the misclassification probability of the \ac{MAP} classifier can be upper bounded as follows:
\begin{equation}
{\bar{P}_{e} }= \sum_{i=1}^L \sum_{\substack{
   j = 1 \\
   j \neq i
  }}^L {\sqrt{p_i p_j}} ~ e^{-K_{ij}},
  \label{multiclass}
\end{equation}
where
%\begin{IEEEeqnarray}{rCl}
%\nonumber
%K_{ij} & =  &   \frac{1}{8}{\bf \Phi}^T\left(\boldsymbol{\mu}_i - \boldsymbol{\mu}_j\right)^T \left[\frac{{\bf \Phi}\left({\bf \Sigma}_i+{\bf \Sigma}_j\right){\bf \Phi}^T + 2 \sigma^2 \mathbf{I}}{2}\right]^{-1}\\
%\nonumber
%&& \cdot {\bf \Phi}\left(\boldsymbol{\mu}_i - \boldsymbol{\mu}_j\right)  \\ 
%&& { + \frac{1}{4}\log\frac{\left(\mathrm{det} \left(\frac{{\bf \Phi}\left({\bf \Sigma}_i+{\bf \Sigma}_j\right){\bf \Phi}^T + 2 \sigma^2 \mathbf{I}}{2}\right)\right)^2}{{\mathrm{det} \left({\bf \Phi}{\bf \Sigma}_i{\bf \Phi}^T + \sigma^2 \mathbf{I}\right)  \mathrm{det} \left({\bf \Phi}{\bf \Sigma}_j{\bf \Phi}^T + \sigma^2 \mathbf{I}\right)}} }. \IEEEeqnarraynumspace
%\label{exp_Bhat}
%\end{IEEEeqnarray}
\begin{IEEEeqnarray}{rCl}
K_{ij} & =  &  {  \frac{1}{4}\log\frac{\left(\mathrm{det} \left(\frac{{\bf \Phi}\left({\bf \Sigma}_i+{\bf \Sigma}_j\right){\bf \Phi}^T + 2 \sigma^2 \mathbf{I}}{2}\right)\right)^2}{{\mathrm{det} \left({\bf \Phi}{\bf \Sigma}_i{\bf \Phi}^T + \sigma^2 \mathbf{I}\right)  \mathrm{det} \left({\bf \Phi}{\bf \Sigma}_j{\bf \Phi}^T + \sigma^2 \mathbf{I}\right)}} }. \IEEEeqnarraynumspace
\label{exp_Bhat}
\end{IEEEeqnarray}
Note that the exponent $K_{ij}$ is a function of the ratio between the volume collectively occupied by {measured} signals belonging to classes $i$ and $j$ and the product of the volumes occupied distinctly by {measured} signals in class $i$ and {measured} signals in class $j$.

The following lemma now provides the low-noise expansion of the upper bound to the probability of error. It defines the probability of error using two quantities: one quantity characterizes the slope of the decay of the upper bound to the misclassification probability (in a $\log \sigma^2$ scale) and the other quantity defines the power offset of the upper bound to the misclassification probability at low-noise levels.

\vspace{0.25cm}

\begin{lemma}
\label{theorem1}
Consider the measurement model in \eqref{s_model} {and the assumptions A.1, A.2 in Section \ref{par:PS}}. Then, in the regime of low noise where $\sigma^2 \to 0$, the upper bound to the probability of misclassification can be expanded as:
\begin{equation}
\bar{P}_{e} = g\left(\sigma^2 \right) ^ { d} + o \left(\left({\sigma^2}\right)^{d}\right),
\label{eq:expansion}
\end{equation}
where
\begin{equation}
d = \min_{\substack{
   i,j \\
   j \neq i
  }} d(i,j)
%\label{diversity_mult_rand}
%\end{equation}
%with
%\begin{equation}
\qv
	d(i,j) = \left(2 r_{ij} - r_i - r_j\right)/4
\label{diver}
\end{equation}
and 
\begin{equation}
g =   \sum_{   (i,j) \in \mathcal{S}_d}  {\sqrt{p_i p_j}} \, 2^{r_{ij}/2}  \left[     \frac{ \sqrt{v_i v_j}}{  v_{ij}  }   \right]^{{1}/{2}}  
\label{gm_mult}
\end{equation}
where $\mathcal{S}_d = \{ (i,j) : i\neq j,  d(i,j) =d \}$. 
\end{lemma}
\vspace{0.25cm}

\begin{IEEEproof}
See Appendix \ref{app_A}.
\end{IEEEproof}
\vspace{0.25cm}

%\begin{lemma}
%\label{theorem2}
%Consider the measurement model in \eqref{s_model} where ${\bf x} \sim \mathcal{N} ({\B{\mu}_i},{\bf \Sigma}_i)$ with probability $p_i$ for $i=1,\ldots, L$. If
%\begin{equation}
%\B{\Phi} (\B{\mu}_i - \B{\mu}_j) \notin \mathrm{Im}( \mathbf{\Phi} (\mathbf{\Sigma}_i + \mathbf{\Sigma}_j) \mathbf{\Phi}^{T}), \forall (i,j), i \neq j,
%  \label{eq:im}
%\end{equation}
%then the upper bound to the probability of misclassification in (\ref{multiclass}) decays exponentially with $1/\sigma^2$ as $\sigma^2 \rightarrow 0$; otherwise, the upper bound to the misclassification probability decays as
%\begin{equation}
%\bar{P}_{e} = a \cdot g \left({\sigma^2} \right) ^ {d} + o \left(\left({\sigma^2}\right)^{d}\right),								
%\end{equation}
%as $\sigma^2 \to 0$, where $a \leq 1$ is a finite constant which depends only on the first term in~(\ref{exp_Bhat}), whereas $d$ and $g$ are given by \eqref{diver} and \eqref{gm_mult}, respectively.
%\end{lemma}
%
%\vspace{0.25cm}
%
%\begin{IEEEproof}
%See Appendix \ref{app_C}.
%\end{IEEEproof}
%
%\vspace{0.25cm}

This lemma leads immediately to the following corollary that provides conditions for $\lim_{\sigma^2 \to 0} \bar{P}_e = 0$ and hence conditions for $\lim_{\sigma^2 \to 0} P_e = 0$.

\vspace{0.25cm}

\begin{corollary}
\label{cor:zero_mean}
Consider the measurement model in \eqref{s_model} {and the assumptions A.1, A.2 in Section \ref{par:PS}}. We have that
\begin{equation}
\exists (i,j), i \neq j: {r_i+r_j} = 2r_{ij} \Rightarrow \lim_{\sigma^2\to 0} \bar{P}_{e} = g >0
\end{equation}
and 
\begin{equation}
{r_i+r_j} <2 r_{ij}, \forall (i,j), i \neq j \Rightarrow \lim_{\sigma^2 \to 0} \bar{P}_{e} = 0.
\end{equation}
\end{corollary}

\vspace{0.25cm}

%\begin{corollary}
%Consider the measurement model in \eqref{s_model} where ${\bf x} \sim \mathcal{N} ({\B{\mu}_i},{\bf \Sigma}_i)$ with probability $p_i$ for $i=1,\ldots, L$. We have that 
%\begin{itemize}
%
%\item if $\B{\Phi} (\B{\mu}_i - \B{\mu}_j) \notin \mathrm{Im}( \mathbf{\Phi} (\mathbf{\Sigma}_i + \mathbf{\Sigma}_j) \mathbf{\Phi}^{T}), \forall (i,j), i \neq j$, then $\lim_{\sigma^2 \to 0} \bar{P}_e = 0$
%
%\item if $\exists (i,j), i \neq j: \B{\Phi} (\B{\mu}_i - \B{\mu}_j) \in \mathrm{Im}( \mathbf{\Phi} (\mathbf{\Sigma}_i + \mathbf{\Sigma}_j) \mathbf{\Phi}^{T})$ and $r_i+r_j=r_{ij}$, then $\lim_{\sigma^2 \to 0} \bar{P}_e = a g>0$
%
%\item if $\forall (i,j), i \neq j$ such that $\B{\Phi} (\B{\mu}_i - \B{\mu}_j) \in \mathrm{Im}( \mathbf{\Phi} (\mathbf{\Sigma}_i + \mathbf{\Sigma}_j) \mathbf{\Phi}^{T})$ it holds ${r_i+r_j} <2 r_{ij}$, then $\lim_{\sigma^2 \to 0} \bar{P}_e = 0$.
%
%\end{itemize}
%\end{corollary}
%\vspace{0.25cm}

The conditions that guarantee that $\lim_{\sigma^2 \to 0} \bar{P}_e = 0$ stem directly from conditions that guarantee $d>0$. The ensuing analysis then concentrates on how to define the effect of the number of random {measurements} or designed {measurements} on the value of the exponent $d$ as a proxy to characterize the phase transition in \eqref{eq:phase_transition}.

\section{Random Measurements}
\label{par:random}

We first consider the simpler problem where the measurement matrix $\mathbf{\Phi}$ is random. In particular, we consider that the measurement matrix is randomly drawn from a left rotation-invariant distribution.\footnote{A random matrix $\mathbf{A}\in \mathbb{R}^{m \times n}$ is said to be (left or right) rotation-invariant if the joint \ac{pdf} of its entries $p(\mathbf{A})$ satisfies $p(\mathbf{\Theta}\mathbf{A})=p(\mathbf{A})$, or $p(\mathbf{A} \mathbf{\Psi})=p(\mathbf{A})$, respectively, for any orthogonal matrix $\mathbf{\Theta}$ or $\mathbf{\Psi}$. A special case of (left and right)  rotation-invariant random matrices is represented by matrices with \ac{i.i.d.}, zero-mean Gaussian entries with fixed variance, which is common in the CS literature~\cite{Candes06a,Donoho06b}.}

We consider the following  problem:

\vspace{0.2cm}
\textit{Determine the minimum number of random measurements needed to guarantee that}
\begin{equation}
\lim_{\sigma^2 \to 0} - \frac{ \log P_e}{\log (1/\sigma^2)} > d_0.
\end{equation}
%\vspace{0.2cm}

%We want to define the solution to the optimization problem:
%
%\begin{equation}
%\begin{aligned}
%& \underset{\mathbf{\Phi} \in S(M,N)}{\text{minimize}}
%& &  M \\
%& \text{subject to}
%& & d(\mathbf{\Phi}) > d_0,
%%& &&  \mathrm{rank} \left(\B{\Phi}\right) \leq M  ,
%\end{aligned}
%\label{eq:PT2classes}
%\end{equation}
%\commentMR{is this problem well formulated in view of the fact that the constraint set S(M,N) and the objective M both depend on M?}
%\commentFR{I think we need to be careful in the definition of the set $S(M,N)$ and the fact that these results hold with probability 1 with respect to the distribution of $\mathbf{\Phi}$. Maybe I can add ``with probability 1'' to the constraint}
%
%\commentFR{think about this formulation}
%
%where $S (M,N)$ is the set of rotation invariant matrices with $M$ rows and $N$ columns and $d(\mathbf{\Phi})$ represents the exponent of the expansion of the upper bound to the misclassification probability in (\ref{eq:expansion}) associated with the measurement matrix $\mathbf{\Phi}$.

The following proposition provides a solution for the case $d_0 = 0$ that leads precisely to the minimum number of measurements for $\lim_{\sigma^2 \to 0} \bar{P}_e = 0$ hence an upper bound on the minimum number of measurements for $\lim_{\sigma^2 \to 0} P_e = 0$.

\vspace{0.25cm}

\begin{proposition}
\label{prop:1}
Consider the measurement model in \eqref{s_model} {and the assumptions A.1-A.4 in Section \ref{par:PS}}. 
%where ${\bf x} \sim \mathcal{N} ({\bf 0},{\bf \Sigma}_i)$ with probability $p_i$, for $i=1,\ldots,L$. Assume also that $\mathrm{rank}(\B{\Sigma}_i) =r_{\B{\Sigma}}<N$ for $i=1,\ldots,L$, and that the images $\mathcal{R}_i$ of the input covariance matrices  are independently drawn from a continuous \ac{pdf} over the Grassmann manifold.\footnote{Note that, if $r_{\mathbf{\Sigma}}=N$, then $\mathcal{R}_i = \mathbb{R}^N, i=1,\ldots,L$, which implies that $\mathrm{Im}(\B{\Phi} \B{\Sigma}_i \B{\Phi}^T)=\mathrm{Im}(\B{\Phi}) $ for all possible values of the measurement kernel $\B{\Phi}$ and, therefore, $d(\mathbf{\Phi})=0$ for all possible values of $\mathbf{\Phi}$.} 
Then, an upper bound on the minimum number of measurements for
\begin{equation}
\lim_{\sigma^2 \to 0} P_e = 0.
\label{eq:phase_prop1}
\end{equation}
is
\begin{equation}
M = r_{\mathbf{\Sigma}}+1.
\label{eq:bound_random}
\end{equation}
\end{proposition}

\vspace{0.25cm}

\begin{IEEEproof}
The proof of this proposition follows immediately from the characterization in Corollary \ref{cor:zero_mean} and from the observation that, with  probability 1 over the distribution of $\mathbf{\Phi}$, $r_i = \min\left\{M,r_{{\bf \Sigma}}\right\}$ and 
$r_{ij} = \min\left\{M, N, 2r_{{\bf \Sigma}}\right\}$.
\end{IEEEproof}
\vspace{0.25cm}

The following proposition provides a generalization of this result from the case $d_0 = 0$ to $d_0 > 0$. %\commentMR{i guess $d_0$ has to be some kind of integer? Please clarify.} \commentFR{Not necessarily. $d$ will take only values that are multiple of $1/4$, actually $1/2$ when we consider covariance images drawn from a continuos pdf on the Grassmann manifold. However, we can take $d_0 \in \mathbb{R}^+$ and the problem is still well defined. On the other hand, if $d_0 \geq R/4$, there exists no $M$ that guarantees $d > d_0$}.

\vspace{0.25cm}

\begin{proposition}
\label{prop:2}
Consider the measurement model in \eqref{s_model} {and the assumptions A.1-A.4 in Section \ref{par:PS}}. 
% where ${\bf x} \sim \mathcal{N} ({\bf 0},{\bf \Sigma}_i)$ with probability $p_i$, for $i=1,\ldots,L$. Assume also that $\mathrm{rank}(\B{\Sigma}_i) =r_{\B{\Sigma}}<N$ for $i=1,\ldots,L$, and that the images $\mathcal{R}_i$ of the input covariance matrices  are independently drawn from a continuous \ac{pdf} over the Grassmann manifold. 
Then, if $d_0 < R/4$, an upper bound on the minimum number of measurements for
\begin{equation}
\lim_{\sigma^2 \to 0} - \frac{ \log P_e}{\log (1/\sigma^2)} > d_0
\label{eq:d_prop2}
\end{equation}
is
\begin{equation}
 M  = \lfloor  2 d_0 +r_{\mathbf{\Sigma}} \rfloor +1.
\end{equation}
%
%On the other hand, if $d_0 \geq R/4$, then
%\begin{equation}
%\lim_{\sigma^2 \to 0} - \frac{ \log P_e}{\log (1/\sigma^2)} \leq d_0 \qv \forall M \in \mathbb{N}.
%\end{equation}
\end{proposition}

\vspace{0.25cm}

\begin{IEEEproof}
The proof of this proposition follows immediately from the characterization of the exponent $d$ in Lemma \ref{theorem1} and from the observation that, with  probability 1 over the distribution of $\mathbf{\Phi}$, $r_i = \min\left\{M,r_{{\bf \Sigma}}\right\}$ and 
$r_{ij} = \min\left\{M, N, 2r_{{\bf \Sigma}}\right\}$.

\end{IEEEproof}
\vspace{0.25cm}

%\commentMR{Please add some remarks about implications of the two propositions above. The remark about the first proposition should elaborate about the interpretation of the minimum number of measurements for the phase transition. The remark about the second proposition should elaborate about such interpretation for the decay.}

We note that the result in Proposition~\ref{prop:1} implies that reliable classification with random measurements is obtained when the signals are embedded into a linear space with dimension strictly greater than the dimension of the spaces spanned by the class conditioned input signals, i.e., $r_{\mathbf{\Sigma}}$; in fact, when this is not the case, the {measured} signals occupy the entire space $\mathbb{R}^{M}$ and, therefore, they are not distinguishable with arbitrarily low misclassification probability when $\sigma^2 \to 0$.

On the other hand, the results in Proposition \ref{prop:2} unveil the interplay between the decay rate of the upper bound to the misclassification probability, the measurements and the geometry of the source. In particular, the results imply that the decay rate scales linearly with the number of measurements up to the maximum decay rate associated with the upper bound in \eqref{multiclass}, i.e., $R/4$, which is achieved when signals are embedded into a linear space with dimension equal to $\dim (\mathcal{R}_{ij}) = \min \{  N, 2r_{\mathbf{\Sigma}} \}$, i.e., the dimension of the sum of any pair of spaces spanned by signals in a given class.

\section{Designed Measurements}
\label{par:2classes}
We now consider the more challenging problem where the measurement matrix $\mathbf{\Phi}$ is designed. In particular, we also want to consider the following problem:

\vspace{0.2cm}
\textit{Determine the minimum number of designed measurements needed to guarantee that}
\begin{equation}
\lim_{\sigma^2 \to 0} - \frac{ \log P_e}{\log (1/\sigma^2)} > d_0.
\end{equation}
%\vspace{0.2cm}

Note once again that by setting $d_0 =0$ one obtains an upper bound to the minimum number of measurements for {$\lim_{\sigma^2\to 0 } P_e = 0$}, thereby guaranteeing a phase transition in the misclassification probability; and by setting $d_0 > 0$ one obtains an upper bound to the minimum number of measurements for $\lim_{\sigma^2 \to 0} - \frac{\log P_e}{\log( 1/\sigma^2)} > d_0 $, thereby guaranteeing a certain decay in the misclassification probability.

We will consider separately the case of two classes and the multiple classes scenario.
 
 \subsection{Two classes}
\label{par:2classes_zeromean}

%The solution of the optimization problem \eqref{eq:PT2classes} for the case of two classes with zero means is given by the following theorem.

The following propositions provide an upper bound to the minimum number of measurements required to drive the misclassification probability to zero  at a rate higher than a given value $d_0$.

\vspace{0.25cm}

\begin{proposition}
\label{prop:3}
Consider the measurement model in \eqref{s_model} {where the assumptions A.1-A.4 in Section \ref{par:PS} are verified and $L=2$}.  
%where ${\bf x} \sim \mathcal{N} ({\bf 0},{\bf \Sigma}_1)$ with probability $p_1$ and ${\bf x} \sim \mathcal{N} ({\bf 0},{\bf \Sigma}_2)$ with probability $p_2 = 1 - p_1$. Assume also that $\mathrm{rank}(\B{\Sigma}_1)=\mathrm{rank}(\B{\Sigma}_2) =r_{\B{\Sigma}}<N$ and that the images $\mathcal{R}_1$ and $\mathcal{R}_2$ are independently drawn from a continuous \ac{pdf} over the Grassmann manifold. 
Then, an upper bound on the minimum number of measurements for
\begin{equation}
\lim_{\sigma^2 \to 0} P_e = 0
\label{eq:phase_prop3}
\end{equation}
is
\begin{equation}
M=1,
\end{equation}
and a possible {measurement} matrix that achieves \eqref{eq:phase_prop3} is obtained by choosing $\B{\Phi} = \B\phi^T$, where $\B\phi \in \mathbb{R}^{N \times 1}$ is a vector in $\mathcal{N}_1$ or $\mathcal{N}_2$ that is not contained in the intersection $\mathcal{N}_1 \cap \mathcal{N}_2$.
\end{proposition}
\vspace{0.25cm}

\begin{IEEEproof}
The proof of this proposition follows immediately from the evaluation of the expansion exponent $d$ of the upper bound \eqref{multiclass}. Namely, when $\B{\Phi} = \B\phi^T$, where $\B\phi \in \mathbb{R}^{N \times 1}$ is a vector in $\mathcal{N}_1$ or $\mathcal{N}_2$ that is not contained in the intersection $\mathcal{N}_1 \cap \mathcal{N}_2$, we obtain $d=(2r_{12} -r_1 - r_2)/4=1/4>0$, which immediately implies \eqref{eq:phase_prop3}. Note also that the existence of the vector $\B\phi$ is guaranteed by the fact that, if $r_{\mathbf{\Sigma}}<N$, then $\mathcal{R}_1 \neq \mathcal{R}_2$ and, therefore, $\mathcal{N}_1 \neq \mathcal{N}_2$.
\end{IEEEproof}
\vspace{0.25cm}

\begin{proposition}
\label{theorem4}
Consider the measurement model in \eqref{s_model} {where the assumptions A.1-A.4 in Section \ref{par:PS} are verified and $L=2$}. Then, if $d_0 <R/4$, an upper bound on the minimum number of measurements for 
\begin{equation}
\lim_{\sigma^2 \to 0} - \frac{ \log P_e}{\log (1/\sigma^2)} > d_0
\label{eq:d_prop4}
\end{equation}
is 
\begin{equation}
M =\lfloor  4 d_0  \rfloor + 1,
\end{equation}
and a {measurement} matrix $\mathbf{\Phi}$ that achieves \eqref{eq:d_prop4} is obtained by choosing arbitrarily $\lfloor  4 d_0  \rfloor + 1$ out of the $R$ rows of matrix
\begin{equation}\label{dmax_mat}
\B{\Phi}_0 = \left[\mathbf{v}_1, \mathbf{v}_2,	\ldots, \mathbf{v}_{n_{\B{\Sigma}}},\mathbf{w}_1,	\mathbf{w}_2,\ldots,\mathbf{w}_{n_{\B{\Sigma}}}\right]^T,
\end{equation}
where the sets $\left[\B{u}_1,\ldots,\B{u}_{n_{12}}\right],~\left[\B{u}_1,\ldots,\B{u}_{n_{12}},\B{v}_1,\ldots,\B{v}_{n_{\B{\Sigma}}}\right]$, $\left[\B{u}_1,\ldots,\B{u}_{n_{12}},\B{w}_1,\ldots,\B{w}_{n_{\B{\Sigma}}}\right]$, $\B{u}_i, \B{v}_i, \B{w}_i  \in \mathbb{R}^N$, constitute an orthonormal basis of the linear spaces $\mathcal{N}_{12}$, $\mathcal{N}_1$ and $\mathcal{N}_2$, respectively, and $n_{12}=[N-2r_{\mathbf{\Sigma}}]^+$, $n_{\B{\Sigma}} = \min \{  N-r_{\mathbf{\Sigma}}, r_{\mathbf{\Sigma}}  \}=R/2$.
%
%On the other hand, if $d_0 \geq R/4$, then
%\begin{equation}
%\lim_{\sigma^2 \to 0} - \frac{ \log P_e}{\log (1/\sigma^2)} \leq d_0 \qv \forall M \in \mathbb{N}.
%\end{equation}
\end{proposition}

\vspace{0.25cm}

\begin{IEEEproof}
See Appendix \ref{app_D}.
\end{IEEEproof}

\vspace{0.25cm}

We can observe that a designed kernel can offer marked improvements over a random one in the low-noise regime. Namely, perfect separation of the {measured} signals can be achieved with a single measurement -- with a random measurement kernel we require $M \geq r_{\B{\Sigma}} +1$ -- and the maximum decay exponent $d$ associated with the upper bound \eqref{multiclass}, i.e., $R/4$, is achieved with $M = R$ -- with a random measurement kernel we require $M= \min\{N,2r_{\mathbf{\Sigma}} \} \geq R$.

We also observe that the kernel design embedded in Proposition \ref{theorem4} relates to previous results in the literature about measurement kernel optimization for the 2-classes classification problem. In particular, for the case of zero-mean classes, it was shown in \cite{Fukunaga90} that the measurement kernel minimizing the Bhattacharyya bound of the misclassification probability for two zero-mean classes is obtained via the eigenvalue decomposition of the matrix $\B{\Sigma}_1^{-1}\B{\Sigma}_2$, where the covariance matrices $\B{\Sigma}_1$ and $\B{\Sigma}_2$ are assumed to be full rank.

A generalization of this construction for the case when $\B{\Sigma}_1$ and $\B{\Sigma}_2$ are not invertible is presented in \cite{Zhang07}. Such kernel design leverages the \ac{GSVD} \cite{Golub96} of the pair of matrices $(\B{\Sigma}_1,\B{\Sigma}_2)$ in order to minimize the corresponding Bhattacharyya upper bound. 
In particular, it is shown that the most discriminant measurements are those corresponding to generalized eigenvectors which lie in the intersections $\mathcal{R}_1 \cap \mathcal{N}_2$ or $\mathcal{R}_2 \cap \mathcal{N}_1$. Then, on recalling that $\mathcal{R}_i = \mathcal{N}_i^\perp$, we can note that the most discriminant measurements are picked from a subspace contained in $\mathcal{N}_1 (\mathcal{N}_2)$ that is also orthogonal to (and therefore, not contained in) $\mathcal{N}_2(\mathcal{N}_1)$. In this sense, the construction described by Proposition~\ref{theorem4} is similar to this result. However, there are significant differences between our results and the results in \cite{Zhang07}. First, our analysis applies to a sensing scenario in \emph{lieu} of feature extraction; so the measurements in (\ref{s_model}) are contaminated by noise whereas the {measurements} in \cite{Zhang07} are not. More importantly, the analysis in \cite{Zhang07} does not offer an explicit characterization of the number of {measurements} needed to guarantee a given misclassification probability performance. On the other hand, our analysis offers sufficient conditions for reliable classification in the low-noise regime and a direct connection between the number of measurements taken on the source signal and the low-noise behavior of the corresponding upper bound to the misclassification probability via the exponent $d$. %Moreover, our construction does not require the computation of the \ac{GSVD} of two matrices, but it simply involves the computation of bases of null spaces of positive semidefinite matrices basis completion of linear spaces.

%\subsection{Two classes: Nonzero-mean}

%It is possible to generalize the results of Propositions \ref{prop:3} and \ref{theorem4} also for the case of two, nonzero-mean classes, i.e., when $\mathbf{x} \sim \mathcal{N} (\B{\mu}_1,\mathbf{\Sigma}_1)$ with probability $p_1$ and $\mathbf{x} \sim \mathcal{N} (\B{\mu}_2,\mathbf{\Sigma}_2)$ with probability $p_2 = 1 - p_1$. In particular, we can observe that, if
%\begin{equation}
%\label{nonzero_design}
%  \B{\mu}_1 - \B{\mu}_2 \notin  \mathcal{R}_{12}.
%\end{equation}
%then, for any finite value $d_{0} \geq 0$, an upper bound to the number of measurements needed to guarantee that $\lim_{\sigma^2 \to 0} - \frac{ \log P_e}{\log (1/\sigma^2)} > d_0$ is $M=1$ and a matrix achieving decay rate is given by
%\begin{equation}
%\B{\Phi} = 
%	\B{\phi}^T     
%\label{phi_nonzero}
%\end{equation}
%where $\B{\phi} $ can be any non-zero vector in $\mathcal{N}_{12}$. In fact, in this case $d(\mathbf{\Phi})$ tends to plus infinity, as the upper bound to the misclassification probability decays exponentially with $1/\sigma^2$.
%
%On the other hand, if
%\begin{equation}
%  \B{\mu}_1 - \B{\mu}_2 \in \mathcal{R}_{12},
%\end{equation}
%then, upper bounds on the number of measurements needed to guarantee \eqref{eq:phase_prop3} and \eqref{eq:d_prop4} are as in Propositions \ref{prop:3} and \ref{theorem4} (see Appendix \ref{app_E}).

\subsection{Multiple classes}
\label{par:PTmulticlass}

%We focus on the case $d_0=0$, thereby determining an upper bound the minimum number of measurements required for the phase transition of the misclassification probability defined in~\eqref{eq:phase_transition}.

The following propositions offer an upper bound to the minimum number of measurements required to drive the misclassification probability to zero, and a procedure to determine an upper bound to the minimum number of measurements required to guarantee that the misclassification probability decays to zero with an exponent higher than a given value $d_0$.

\vspace{0.25cm}

\begin{proposition}
\label{theo:PTmulticlass}
Consider the measurement model in \eqref{s_model} {and the assumptions A.1-A.4 in Section \ref{par:PS}}.  
Then, an upper bound on the minimum number of measurements for
\begin{equation}
\lim_{\sigma^2 \to 0} P_e=0
\label{eq:PT_multiple}
\end{equation}
is
\begin{equation}
M  = \min \{ L-1, r_{\B{\Sigma}}+1   \}.
\label{eq:UB_PT_multiclass}
\end{equation}
Moreover, a {measurement} matrix $\mathbf{\Phi}$ that achieves \eqref{eq:PT_multiple} is obtained as follows: let $\mathbf{N}_i$ be a matrix that contains a basis for the null space $\mathcal{N}_i$. Then, the $M= \min \{ L-1, r_{\B{\Sigma}}+1   \}$ rows of the matrix $\mathbf{\Phi}$ are obtained by randomly picking one row from each of the matrices $\mathbf{N}_{\pi(1)}^T,\ldots, \mathbf{N}_{\pi( \min \{ L-1, r_{\B{\Sigma}}+1   \})}^T$, where $\pi(\cdot)$ is any permutation function of the integers $1,\ldots,L$.
%
%
%
% by picking each row from a random basis of $\min \{ L-1, r_{\B{\Sigma}}+1   \}$ different null spaces among $\mathcal{N}_1,\ldots, \mathcal{N}_L$. \commentFR{explain better}
\end{proposition}

\vspace{0.25cm}

\begin{IEEEproof}
See Appendix~\ref{app:design_multiclass}.
\end{IEEEproof}
\vspace{0.25cm}

Note that the characterization embodied in Proposition \ref{theo:PTmulticlass} is obtained by taking the measurement matrix to belong to a certain restricted subset of $\mathbb{R}^{M \times N}$ rather than the entire $\mathbb{R}^{M \times N}$. %Therefore, the upper bound on the number of measurements required to guarantee (\ref{eq:PT_multiple}) associated with the solution of (\ref{eq:PT2classes}) under this restricted measurement set can be looser than that associated with the set $\mathbb{R}^{M \times N}$.

The choice of such subset of $\mathbb{R}^{M \times N}$ is inspired by our characterization pertaining to the two-class problem embodied in Propositions \ref{prop:3} and \ref{theorem4}. Namely, let $\mathbf{N}_i \in \mathbb{R}^{N \times (N-r_{\B{\Sigma}})}$ be a matrix that contains a basis for the null space $\mathcal{N}_i$ and let $\mathbf{N} = [\mathbf{N}_1, \ldots, \mathbf{N}_L]$ be a matrix that contains the concatenation of the bases for all the null spaces $\mathcal{N}_1,\ldots,\mathcal{N}_L$. Then, we take the {measurement} matrix to consist of $M$ rows of $\mathbf{N}^T$ rather than $M$ arbitrary vectors from $\mathbb{R}^N$.

Note also that the result embodied in Proposition~\ref{theo:PTmulticlass} -- which is shown to be very sharp both with synthetic data and real data simulations -- provides a fundamental insight in the role of measurement design in comparison with random measurement kernels in the discrimination of subspaces. In particular, we can clearly identify two operational regimes that depend on the relationship between two fundamental geometrical parameters describing the source: the number of classes and the dimension of the linear subspaces associated to the different classes. 
\begin{itemize}
\item When, the number of classes in the source is lower than or equal to the dimension of the spaces spanned by signals in each class, the designed {measurement} matrix is such that we take one measurement from $L-1$ out of the $L$ null spaces $\mathcal{N}_i, i=1,\ldots,L$. In this sense, the construction that achieves the upper bound implements a \emph{one-vs-all} approach, where each measurement is able to perfectly detect the presence of signals coming from a specific class against signals from all the remaining classes. Note that in this regime, proper design of the measurement kernel can provide a dramatic performance advantage with respect to random measurements, as it can guarantee that the misclassification probability approaches zero,  in the low-noise regime, even when random measurements yield an error floor.

\item On the other hand, when the number of classes is larger than the dimension spanned by signals in a given class, (more precisely, when $L > r_{\B{\Sigma}}+1$), then $r_{\B{\Sigma}}+1$ measurements are sufficient to drive to zero the misclassification probability in the low-noise regime. In this case, the designed measurement kernel obtains the same performance of random measurements in terms of phase transition of upper bounds to the misclassification probability. However, properly designing the measurement kernel can have an impact on the value of the error floor or the speed of the decay of the misclassification probability with $1/\sigma^2$. 
\end{itemize}

\vspace{0.25cm}

\begin{proposition}
\label{prop:6}
Consider the measurement model in \eqref{s_model}{and the assumptions A.1-A.4 in Section \ref{par:PS}}. Then, if $d_0<R/4$, an upper bound on the minimum number of measurements for
\begin{equation}
\lim_{\sigma^2 \to 0} - \frac{ \log P_e}{\log (1/\sigma^2)} > d_0
\label{eq:d_prop6}
\end{equation}
is given by the solution to the integer programming problem
\begin{equation}
\begin{aligned}
& \underset{(M_1,\ldots, M_L ) \in \mathbb{N}^L}{\text{minimize}}
& &  M  = \sum_{i=1}^L M_i \\
& \text{subject to:}
& & M_i \leq N-r_{\mathbf{\Sigma}} ,  \forall i \\
& &&   f(M,M_i, M_j) -2(M-2r_{\mathbf{\Sigma}}) > d_0, \forall i \neq j  \\
& &&   f(M,M_i, M_j) -2(M_i-r_{\mathbf{\Sigma}}) > d_0, \forall i \neq j  \\
& &&   f(M,M_i, M_j) -2(M_j-r_{\mathbf{\Sigma}}) > d_0, \forall i \neq j  \\
& &&   f(M,M_i, M_j)  > d_0, \forall i \neq j  ,
\end{aligned}
\label{eq:IntProg}
\end{equation}
where $f(M,M_i,M_j)=\max \{  M- r_{\B{\Sigma}}, M_i\} + \max \{  M- r_{\B{\Sigma}}, M_j\}$. 
%
%On the other hand, if $d_0 \geq R/4$, then
%\begin{equation}
%\lim_{\sigma^2 \to 0} - \frac{ \log P_e}{\log (1/\sigma^2)} \leq d_0 \qv \forall M \in \mathbb{N}.
%\end{equation}
%
%
\end{proposition}

\vspace{0.2cm}

\begin{IEEEproof}
Note that $R/4$ is the maximum decay exponent $d$ associated with the upper bound in \eqref{multiclass}, and note also that, if $d_0 <R/4$, a sufficient condition for \eqref{eq:d_prop6} is given by $d(i,j) >d_0$, for all $(i,j)$, $i \neq j$. Then, the proposed upper bound follows from taking the measurement matrix to belong to the same restricted subset considered in Proposition~\ref{theo:PTmulticlass}. In this case, on denoting by $M_i$ the number of measurements in $\mathbf{\Phi}$ that are also columns of $\mathbf{N}_i$, so that $M = \sum_{i=1}^L M_i$, we can write\footnote{The details are provided in Appendix~\ref{app:design_multiclass}.}
\begin{IEEEeqnarray}{rCl}
\nonumber
%d(i,j) & = & \max\{  M-r_{\B{\Sigma}} , M_i \} +  \max\{  M-r_{\B{\Sigma}} , M_j \} \\
 d(i,j) & = &f(M,M_i,M_j) \\
 && - 2 \max\{  M- 2 r_{\B{\Sigma}} , M_i-r_{\B{\Sigma}} , M_j-r_{\B{\Sigma}} , 0 \},
\end{IEEEeqnarray}
which leads to the formulation of the problem \eqref{eq:IntProg}.
\end{IEEEproof}

\vspace{0.25cm}

Note that, although a general closed-form solution to the optimization problem in \eqref{eq:IntProg} is difficult to provide, our formulation allows to drastically reduce the number of (integer) optimization variables, which is now equal to the number of classes $L$. Moreover, the integer programming problem in \eqref{eq:IntProg} involves a linear objective function and constraints that are expressed via linear functions combined via the max function, thus allowing the use of efficient numerical methods for its solution.

It is also important to emphasize the differences between the result in Proposition~\ref{theo:PTmulticlass} and other results in the literature. The result in \eqref{eq:UB_PT_multiclass} is reminiscent of a result associated to multiclass \ac{LDA}, that involves the extraction of $L-1$ linear features from the data using the \ac{LDA} rule \cite{duda00}. However, such \ac{LDA} construction does not provide conditions on the number of measurements needed for reliable classification. Moreover, in contrast with the analysis here proposed, \ac{LDA} approaches are usually applied to the nonzero-mean classes scenario rather than the zero-mean case considered here. In fact, \ac{LDA} methods are shown to be ineffective in the case of zero-mean classes, due to the measurement kernel construction approach that is based on the computation of the \ac{GSVD} of inter-class and intra-class scatter matrices, where the first one is a function of the class means.

A modified version of \ac{LDA} which can cope also with zero-mean classes has been presented in \cite{Zhang07}. Such method is based on recasting a multiclass classification problem into a binary pattern classification problem. However, in this case the measurement kernel $\mathbf{\Phi}$ is not determined on the basis of the statistical description of the classes, but rather it is derived via a non-parametric approach, which involves the computation of scatter matrices from labeled training samples. 
{In particular, on denoting by $\mathbf{\Sigma}\sub{b}$ the between-class scatter matrix and by $\mathbf{\Sigma}\sub{w}$ the within-class scatter matrix, measurements are designed in order to maximize the objective function
\begin{equation}
J(\mathbf{\Phi})  = \mathrm{tr}\left(  (\mathbf{\Phi}  \mathbf{\Sigma}\sub{w} \mathbf{\Phi}^T)^{-1}  (\mathbf{\Phi}  \mathbf{\Sigma}\sub{b} \mathbf{\Phi}^T) \right),
\end{equation}
leading to measurement designs that are associated with the generalized eigenvectors corresponding to the largest generalized eigenvalues of $(\mathbf{\Sigma}\sub{b}, \mathbf{\Sigma}\sub{w})$.}
In addition, in this case, conditions on number of measurements needed for reliable classification are not available in general. 
%In fact, it is possible to show that the misclassification probability achieved by $L-1$ projections designed according to such an \ac{LDA} approach is identical to that obtained in the original domain (thus approaches zero in the low-noise regime) under the homoscedastic assumption, that entails that $\B{\Sigma}_i=\B{\Sigma}$ for $i=1, \ldots , L$ \cite{Nenadic07}. 
%
%Similarly to \ac{LDA}, also $L-1$ measurements taken according to the \ac{IDA} approach guarantee zero classification error in the low-noise regime under the homoscedastic assumption\footnote{In fact, when the homoscedastic assumption is verified, \ac{IDA} and \ac{LDA} are equivalent.} \cite{Nenadic07}, or under certain assumptions on the geometrical description of the classes \cite[Section 3.5]{Nenadic07}. 

\section{Numerical Results}
\label{par:num_res}

We now show how our theory aligns with practice, both for synthetic data and real data associated with a video segmentation application {and with a face recognition application}. We also show how our upper bound on the minimum number of measurements required for the phase transition compares to those associated with state-of-the-art measurement designs such as \ac{IDA} methods \cite{Nenadic07} and methods based on the maximization of Shannon mutual information and quadratic R\'enyi entropy \cite{Chen12}.

\subsection{Synthetic data}
\label{par:ResSynth}

{We first consider experiments with synthetic data by concentrating on two examples that reflect the two regimes embodied in Proposition \ref{theo:PTmulticlass}. In the first example, the data is generated by a mixture of $L=11$ Gaussian distributions with dimension $N=64$, with probability $p_i=1/11$, for $i=1,\ldots,11$. The input covariance matrices have all rank $r_{\mathbf{\Sigma}}=14$, and their images are drawn uniformly at random from the Grassmann manifold of $14$-dimensional spaces in $\mathbb{R}^{64}$. %We also assume $\boldsymbol{\mu}_i=\mathbf{0}$ for $i=1,2,3$.
}

Figure \ref{fig:1Bhat} reports the upper bound to the misclassification probability and the true misclassification probability, respectively, vs $1/\sigma^2$ both for random kernel designs and measurement designs that obey the construction embodied in Proposition \ref{theo:PTmulticlass}.\footnote{Note that the construction embodied in Proposition \ref{theo:PTmulticlass} is shown to achieve the low-noise phase transition with a number of measurements equal to  \eqref{eq:UB_PT_multiclass}.} The {measurement} kernels are also normalized such that $\mathrm{tr}(\mathbf{\Phi}^T\mathbf{\Phi}) \leq M$.

Note that theoretical results are aligned with experimental results in the sense that both theory and practice suggest that the low-noise phase transition occurs with $M \geq L-1=10$ for designed kernels and $M \geq r_{\mathbf{\Sigma}}+1=15$ for random kernels. This is observed from Fig. \ref{fig:1Bhat}, suggesting that our analysis is sharp.

%%%%%%

%\begin{figure}[tb]
%\begin{center}
%\subfigure[Upper bound $\bar{P}_e$]{\includegraphics[width=0.42\textwidth]{Fig1Bhat.eps}}
%\subfigure[Misclassification probability $P_e$]{\includegraphics[width=0.42\textwidth]{Fig1Perr.eps}}
%\end{center}
%\caption{Upper bound and true misclassification probability vs. $1/\sigma^2$. $N=6$, $L=3$, $r_{\mathbf{\Sigma}}=4$. Random measurement kernels (dashed lines) and designed kernels (solid lines).}
%\label{fig:1Bhat}
%\end{figure}

{
\begin{figure}[tb]
\begin{center}
\includegraphics[width=0.42\textwidth]{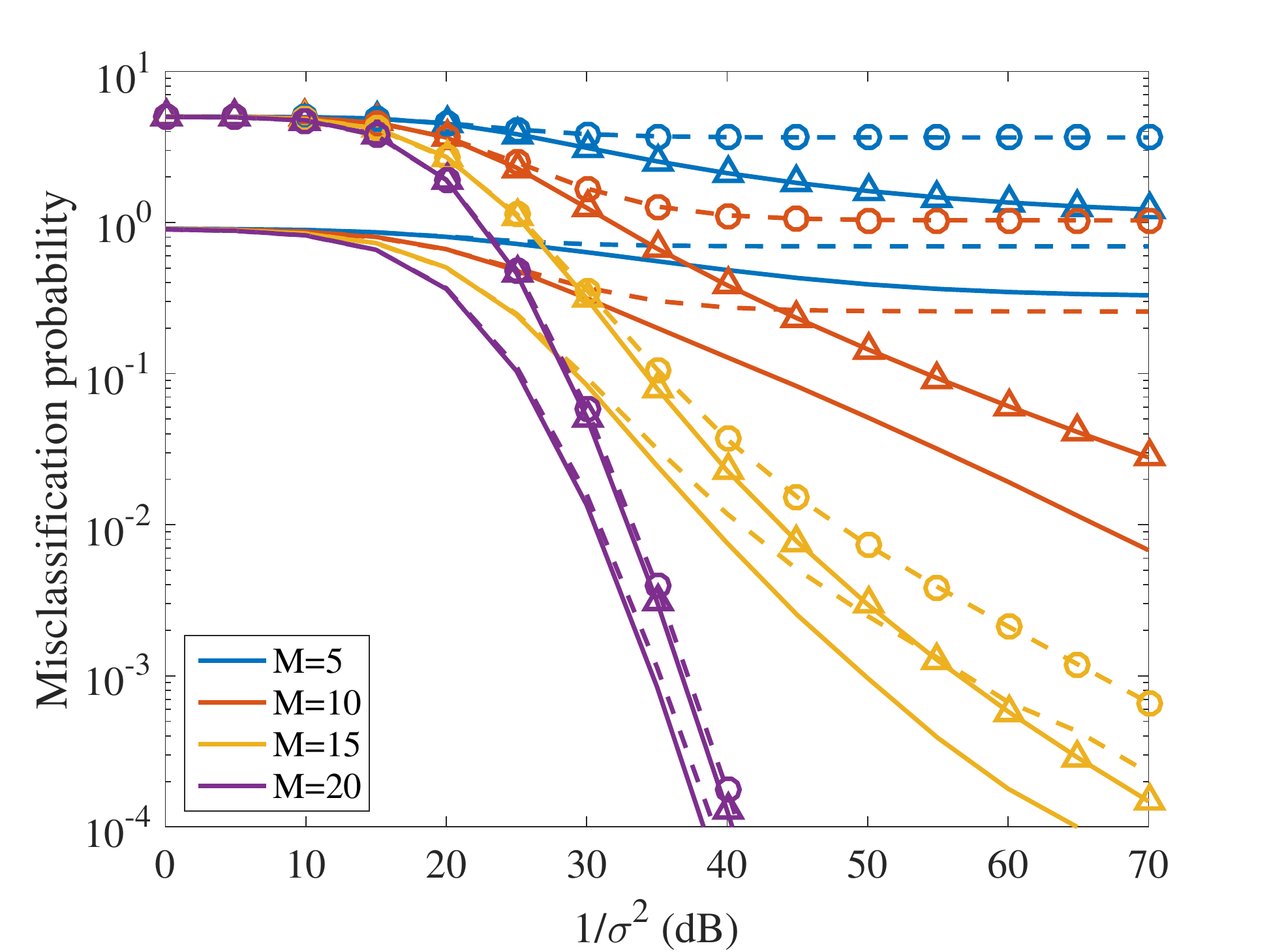}
\end{center}
\caption{Upper bound and true misclassification probability vs. $1/\sigma^2$. $N=64$, $L=11$, $r_{\mathbf{\Sigma}}=14$. True misclassification probability with random measurement kernels (dashed lines) and designed kernels (solid lines). Upper bound to the misclassification probability with random measurement kernels (dashed lines with circles) and designed kernels (solid lines with triangles).}
\label{fig:1Bhat}
\end{figure}
}
{
In the second example, the data is drawn from a mixture of $L=12$ Gaussian distributions with dimension $N=64$, with probability $p_i=1/12$ for $i=1, \ldots,12$. The input covariance matrices have all rank $r_{\mathbf{\Sigma}}=9$, and their images are drawn uniformly at random from the Grassmann manifold of $9$-dimensional spaces in $\mathbb{R}^{64}$.% We also assume $\boldsymbol{\mu}_i=\mathbf{0}$ for $i=1,\ldots,6$.

Figure \ref{fig:2Bhat} showcases the upper bound to the misclassification probability and the true misclassification probability, respectively, vs $1/\sigma^2$ both for random kernel designs and measurement designs that obey the construction embodied in Proposition \ref{theo:PTmulticlass}. It is evident -- as predicted by Proposition \ref{theo:PTmulticlass} -- that both random and designed kernels achieve a low-noise phase transition in the upper bound to the misclassification probability with $M \geq r_{\mathbf{\Sigma}}+1=10$. However, designed kernels offer a lower misclassification probability than random kernels for finite noise levels. It is also evident by comparing the true misclassification probability values and the upper bounds in Fig. \ref{fig:2Bhat} that our analysis is sharp.
}

%\begin{figure}[tb]
%\begin{center}
%\subfigure[Upper bound $\bar{P}_e$]{\includegraphics[width=0.42\textwidth]{Fig2Bhat.eps}}
%\subfigure[Misclassification probability $P_e$]{\includegraphics[width=0.42\textwidth]{Fig2Perr.eps}}
%\end{center}
%\caption{Upper bound and true misclassification probability vs. $1/\sigma^2$. $N=6$, $L=6$, $r_{\mathbf{\Sigma}}=3$. Random measurement kernels (dashed lines) and designed kernels (solid lines).}
%\label{fig:2Bhat}
%\end{figure}

{
\begin{figure}[tb]
\begin{center}
\includegraphics[width=0.42\textwidth]{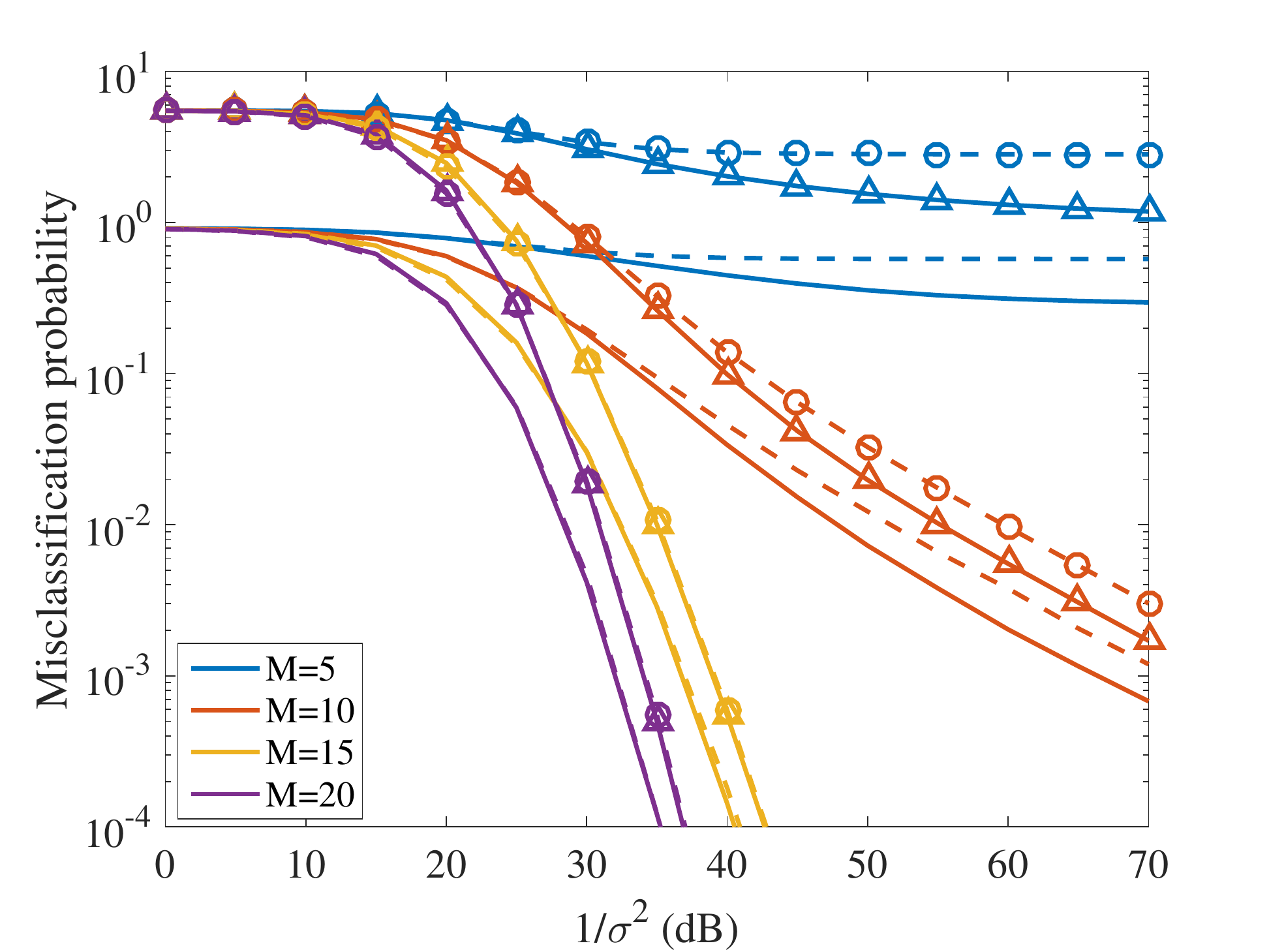}
\end{center}
\caption{Upper bound and true misclassification probability vs. $1/\sigma^2$. $N=64$, $L=12$, $r_{\mathbf{\Sigma}}=9$. True misclassification probability with random measurement kernels (dashed lines) and designed kernels (solid lines). Upper bound to the misclassification probability with random measurement kernels (dashed lines with circles) and designed kernels (solid lines with triangles).}
\label{fig:2Bhat}
\end{figure}
}

Our upper bound to the minimum number of measurements required for the phase transition of the misclassification probability relied on a specific construction. It is therefore relevant to examine how such a bound compares to the number of measurements required for the phase transition associated with state-of-the-art kernel designs. To that end, we consider three state-of-the-art {measurement} kernel designs applied to the two previous examples: these are the \ac{IDA} method in \cite{Nenadic07} and methods based on the maximization of Shannon mutual information (MI) and R\'enyi quadratic entropy \cite{Chen12}, respectively. 

Table~\ref{table:1} reports the minimum number of measurements needed by such methods in order to drive to zero the numerically simulated misclassification probability, as well as the theoretical predictions derived in the previous sections for both random and designed kernel. %\footnote{Actual values of $P_e$ vs. $1/\sigma^2$ are not reported for space reasons.} 
It is interesting to see that the bound embodied in Proposition \ref{theo:PTmulticlass} predicts very well the behavior of state-of-the-art kernel design methods. This means that our bound can be used to gauge a suitable number of measurements to be used in state-of-the-art kernel design approaches.

%%%%%%%%%%%%
%We now apply different state-of-the-art projection design methods to measure synthetic data generated according to the sources in case i) and case ii). In particular, we consider projection kernels designed via the \ac{IDA} method in \cite{Nenadic07} and via the maximization of Shannon mutual information (MI) and R\'enyi quadratic entropy \cite{Chen12}. In Table~\ref{table:1} are reported the minimum number of measurements needed by such methods in order to drive to zero the numerically simulated misclassification probability, as well as the theoretical predictions derived in the previous sections for both random and designed kernel.\footnote{Actual values of $P_e$ vs. $1/\sigma^2$ are not reported for space reasons.} Then, It is possible to observe that the bound in \eqref{eq:UB_PT_multiclass} predicts well the behavior of state-of-the-art kernel design methods. 

\begin{table}
\caption{Minimum number of measurements $M$ required to achieve the low-noise phase transition of the misclassification probability.}% for different projection kernels. Synthetic data.}
\begin{center}
\begin{tabular}{l|c|c|c|c|c}
    & random  & design \eqref{eq:UB_PT_multiclass} & IDA & MI & R\'enyi \\
    \hline
 case i) &  15  & 10  & 10 & 10 & 10  \\
 case ii) &  10  & 10  & 10 & 10 & 10
\end{tabular}
\end{center}
\label{table:1}
\end{table}

\subsection{Real data: Motion segmentation}
\label{par:Hopkins}

We now consider experiments with real  data by concentrating on a motion segmentation application, where the goal is to segment a video in multiple rigidly moving objects. Such application involves the extraction of feature points from the video whose position is tracked over different frames. Then, motion segmentation {aims at partitioning pixels extracted from different frames into spatiotemporal regions. In particular, feature point are clustered into} different groups, each corresponding to a given motion~\cite{vidalTutorial}. The data {to be processed by the clustering algorithm} is obtained by stacking the coordinate values associated to a given feature point corresponding to different frames. {For a detailed description of how clustering data are obtained from feature points coordinates, please refer to~\cite{vidalTutorial}.}

We use the Hopkins 155 motion segmentation dataset \cite{Tron2007}, which  consists of video sequences with two or three motions in each video. {Each video of two motions consists of 30 frames, whereas each video of three motions consists of 29 frames.} In particular the results reported in this section are obtained by considering the video with three motions in the dataset having the largest number of samples for each motion/class\footnote{Denoted as ``1RT2RCR'' in the dataset.}, {namely, 142 samples for class 1, 114 samples for class 2 and 236 samples for class 3.}

We consider in particular a supervised learning approach, in which {$50 \%$ or $30 \%$} of the vectors corresponding to features points are manually labeled, whereas the remaining points are classified automatically, starting from the observation of noisy measurements, where the noise variance is set to $\sigma^2 = -60$\,dB. The manually labeled points represent labeled training samples from which the input signal parameters $p_i, \mathbf{\Sigma}_i$, $i=1,\ldots,L$ are inferred using \ac{ML} estimators.%\footnote{We assume $\boldsymbol{\mu}_i=\mathbf{0}$ for $i=1,\ldots,L$.}.

As described in \cite{vidalTutorial,Tomasi1992,Boult1991}, features points trajectories belonging to a given motion can be shown to lie on approximately three dimensional affine spaces or four dimensional linear spaces. In fact, the covariance matrices obtained from the training samples present only two dominant principal components, as demonstrated by the magnitudes of eigenvalues of the input covariance matrices reported in Table~\ref{table:Eig}. Then, based on the results presented in Propositions~\ref{prop:1} and \ref{theo:PTmulticlass}, we can expect that at least 3 random measurements and 2 designed measurements are needed for reliable classification, respectively.

\begin{table}
\caption{Eigenvalues of the input covariance matrices obtained from training samples from the video ``1RT2RCR'' via the \ac{ML} estimator. Largest five eigenvalues for each class.}
\begin{center}
\begin{tabular}{l||c|c|c|c|c}
$\mathbf{\Sigma}_1$ & 9.6284 & 2.2694 & 0.0194 & 0.0061 & 0.003 \\
$\mathbf{\Sigma}_2$ & 3.1756 &  0.7410 & 0.0267 &  0.0022 & 0.000 \\
$\mathbf{\Sigma}_3$ &11.2797 & 5.9315 & 0.0672 & 0.0004 & 0.000
\end{tabular}
\end{center}
\label{table:Eig}
\end{table}

\begin{figure}[tb]
\begin{center}
\subfigure[$50 \%$ training samples]{\includegraphics[width=0.42\textwidth]{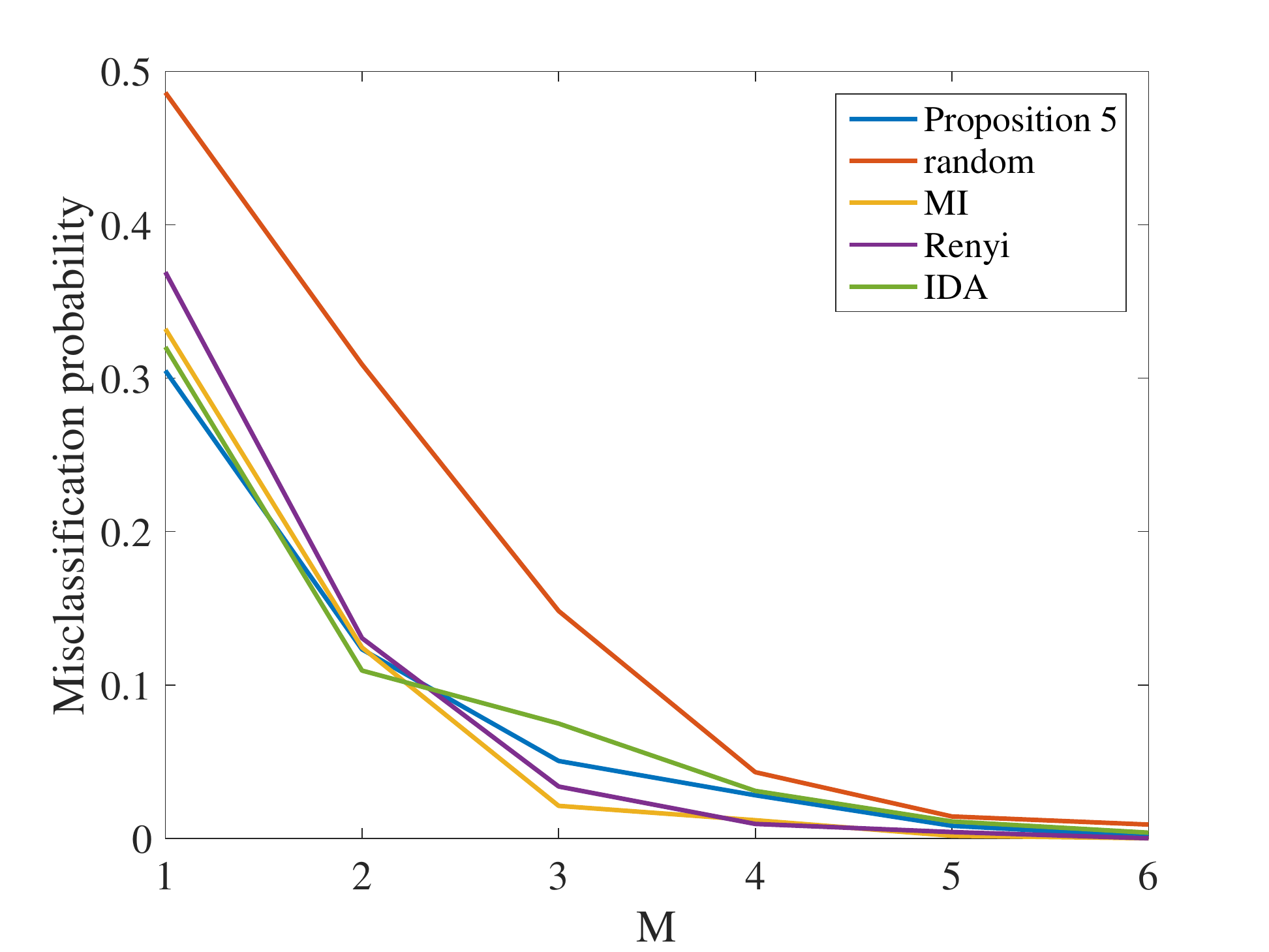}}
\subfigure[$30 \%$ training samples]{\includegraphics[width=0.42\textwidth]{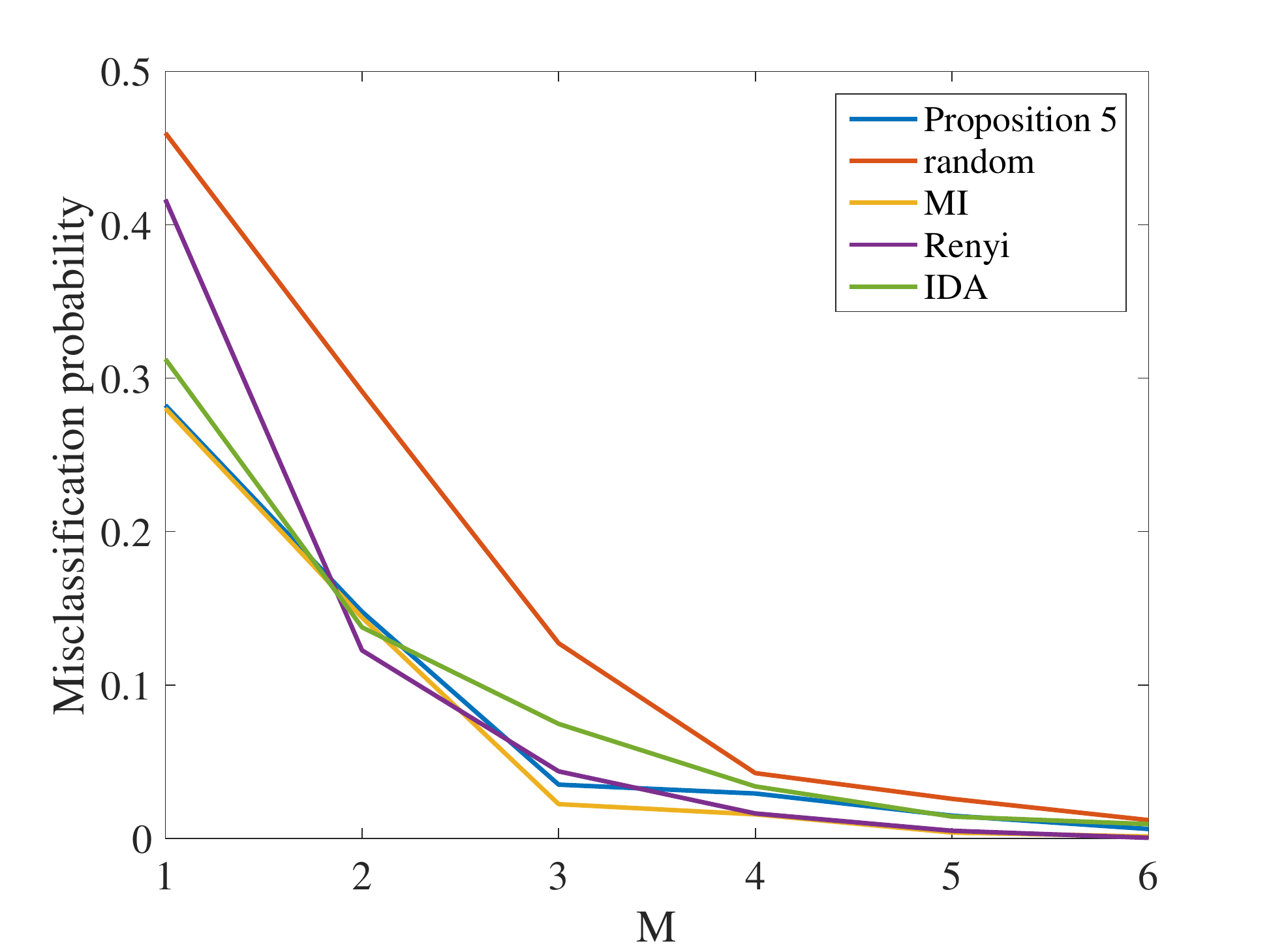}}
%\subfigure[$10 \%$ training samples]{\includegraphics[width=0.42\textwidth]{PerrVsM_01train.eps}}
\end{center}
\caption{Misclassification probability vs. $M$. Hopkins 155, 1RT2RCR dataset. $50 \%$ or $30 \%$ of the samples are manually labeled and used for training.}
\label{fig:PerrVsM}
\end{figure}

Figures~\ref{fig:PerrVsM} (a) and (b) report the misclassification probability vs the number of measurements for random kernels, kernels designed via the construction embodied in Proposition~\ref{theo:PTmulticlass}, and the designs in \cite{Nenadic07,Chen12}. {In particular, in view of the fact that the analysis is conducted for the scenario where the \ac{MAP} classifier is provided with the true model parameters, our results consider both the scenario where a significant number of training samples ($50\%$) is used to learn the underlying models and a scenario where a lower number of training samples ($30\%$) is used to derive the models in order to assess the robustness of the theoretical insights agains model mismatch.} Note that now the misclassification probability does not exhibit a perfect phase transition in view of the fact that the data covariance matrices are not low-rank anymore but rather approximately low-rank, {and due to the mismatch between the model inferred from training data and the actual test data}. However, one can still conclude that our theoretical results align with practical ones, since they can unveil the number of measurements required for the misclassification probability to be below a certain low value.

In particular, Table~\ref{table:Real} reports the minimum number of measurements required by the random and designed kernels to achieve a misclassification probability below $15\%$, $10\%$ and $5\%$, {for both cases when $50 \%$ and $30\%$ of the vectors in the dataset are used as training samples}. It can be observed that our characterization of the upper bound to the number of measurements required for the phase transition matches well the number of measurements required to achieve a low misclassification probability in \ac{IDA} and methods based on the maximization of Shannon mutual information and R\'enyi quadratic entropy, {in both scenarios where $50 \%$ and $30\%$ of the vectors in the dataset are used as training samples.}

\begin{table}
\caption{Minimum number of measurements $M$ required to achieve a given value of the misclassification probability. Hopkins 155 dataset.}
\begin{center}
\begin{tabular}{l|c|c|c|c|c}
    & random & Prop. \ref{theo:PTmulticlass} & IDA & MI & R\'enyi \\
    \hline
    \multicolumn{6}{c}{$50 \%$ training samples}\\
    \hline
$P_{e} < 15 \%$ &  3 & 2  & 2  & 2 & 2   \\
$P_{e} < 10 \%$ &  4 & 3 & 3  & 3 & 3 \\
$P_{e} < 05 \%$ &  4 & 4 & 4  & 3 & 3 \\
    \hline
    \multicolumn{6}{c}{$30 \%$ training samples}\\
    \hline
$P_{e} < 15 \%$ &  3 & 2  & 2  & 2 & 2   \\
$P_{e} < 10 \%$ &  4 & 3 & 3  & 3 & 3 \\
$P_{e} < 05 \%$ &  4 & 3 & 4  & 3 & 3 
\end{tabular}
\end{center}
\label{table:Real}
\end{table}

%\begin{table}
%\caption{Minimum number of measurements $M$ required to achieve a given value of the misclassification probability. Hopkins 155 dataset.}
%\begin{center}
%\begin{tabular}{l|c|c|c|c|c}
%    \multicolumn{6}{c}{$50 \%$ training samples}\\
%    \hline
%    & random & Prop. \ref{theo:PTmulticlass} & IDA & MI & R\'enyi \\
%    \hline
%$P_{e} < 15 \%$ &  3 & 2  & 2  & 2 & 2   \\
%$P_{e} < 10 \%$ &  4 & 3 & 3  & 3 & 3 \\
%$P_{e} < 05 \%$ &  4 & 4 & 4  & 3 & 3 \\
%    \hline
%    \multicolumn{6}{c}{$30 \%$ training samples}\\
%    \hline
%    & random & Prop. \ref{theo:PTmulticlass} & IDA & MI & R\'enyi \\
%    \hline
%$P_{e} < 15 \%$ &  3 & 2  & 2  & 2 & 2   \\
%$P_{e} < 10 \%$ &  4 & 3 & 3  & 3 & 3 \\
%$P_{e} < 05 \%$ &  4 & 3 & 4  & 3 & 3 
%\end{tabular}
%\end{center}
%\label{table:Real}
%\end{table}

{\subsection{Real data: Face recognition}
\label{par:Faces}

%\commentMR{Numerical results for face recognition are obtained by considering $L=5$ faces from the YaleB dataset, and by using different amounts of training samples. The results are quite noisy, due to the fact that the number of samples for each class are not too many. Moreover, averaging over different choices of which samples are used for training and which are used for test is slightly time consuming, due to the larger dimension of the optimization problems solved for mutual information, Renyi and IDA measurement designs (these curves are obtained with a 10 hours simulation each one). I also think that our proposed design does not perform very well in this case since it is based on the assumption that classes are exactly low-rank (measeumrents are taken on the null spaces associated to different classes), whereas real data have a slower eigenvalue decay with respect to Hopkins.}

We now consider a different real-word, compressive classification application. In particular, we consider a face recognition problem where the orientation of faces associated to different individuals relative to the camera remains fixed, but the illumination conditions vary. On assuming that faces are approximately convex and that reflect light according to Lambert's law, it is possible to show that the set of images of a same individual under different illuminations lies approximately on a 9-dimensional linear subspace \cite{Basri03}. Therefore, face recognition from linear measurements extracted from such images can be performed via subspace classification.

In this section, we show classification results using cropped images from the Extended Yale Face Database B~\cite{Georghiades01}. In particular, we consider $16 \times 16$ images of $L=5$ different individuals from the $38$ available in the dataset. For each individual, $63$ images corresponding to $63$ different illumination conditions are considered. 

As for the video motion segmentation application described in Section~\ref{par:Hopkins}, classification is performed via the \ac{MAP} classifier \eqref{class}, where we assume Gaussian distribution for each class and the parameters $p_i, \mathbf{\Sigma}_i$ are obtained via \ac{ML} estimators by using $50 \%$ or $30 \%$ of the available images as training samples. Moreover, we set the noise variance to $\sigma^2=-60$ dB.

\begin{figure}[tb]
\begin{center}
\includegraphics[width=0.42\textwidth]{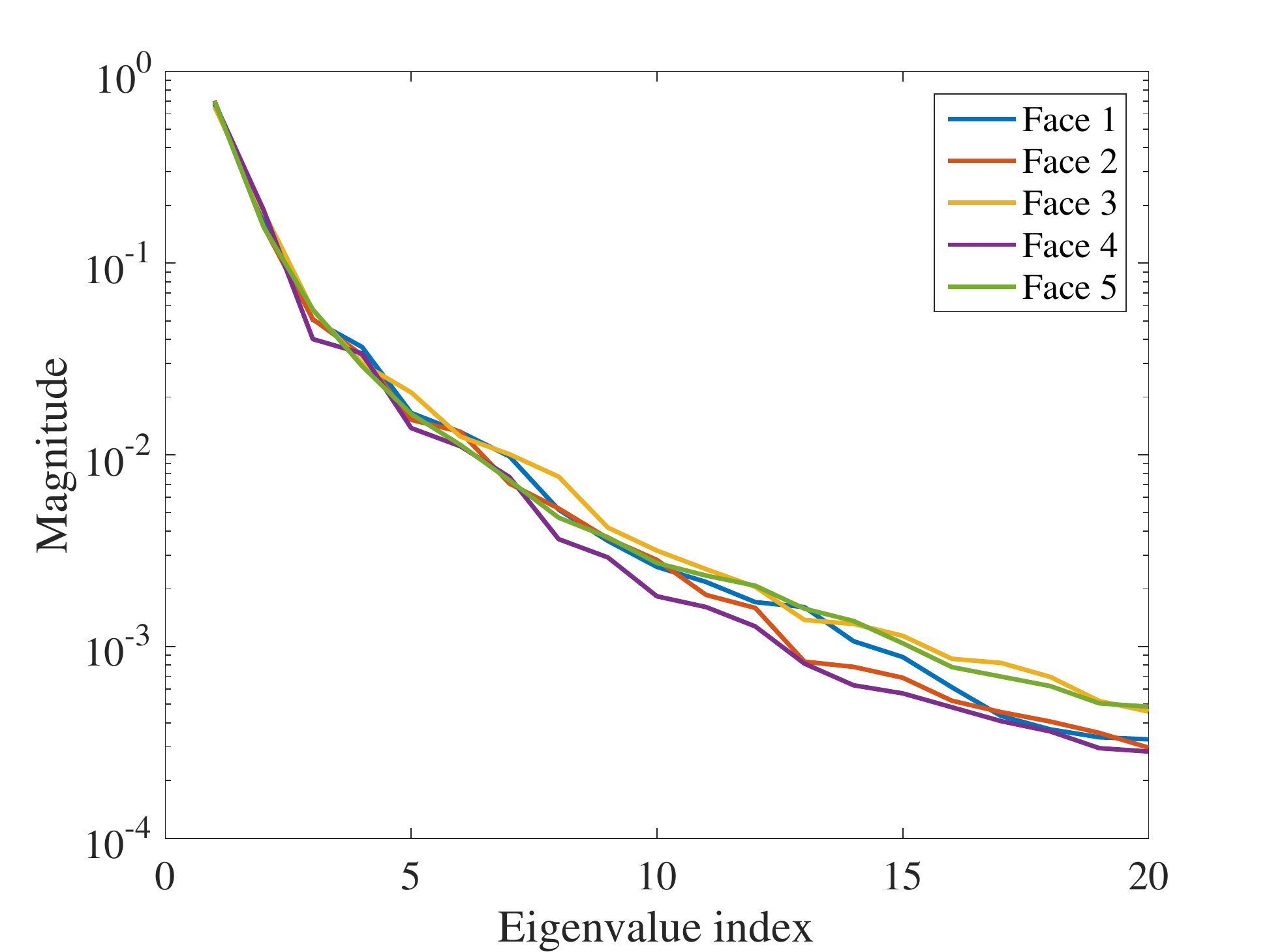}
\end{center}
\caption{Largest $20$ eigenvalues of the covariance matrices associated to the first $L=5$ classes in the Extended Yale Face Database B. The covariance matrices are obtained from training samples via the \ac{ML} estimator.}
\label{fig:EigFaces}
\end{figure}

In contrast with the case of the Hopkins 155 dataset, samples in the Extended Yale Face Database B are described via an approximately low-rank model which is characterized by a slower decay of the eigenvalues of the corresponding covariance matrices, as reported in Fig. \ref{fig:EigFaces}. In this sense, experimental results for this dataset represent a way to test the predictions provided by our analysis also for a scenario which departs further from the assumption of signals lying on a union of low-dimensional subspaces.

\begin{figure}[tb]
\begin{center}
\subfigure[$50 \%$ training samples]{\includegraphics[width=0.42\textwidth]{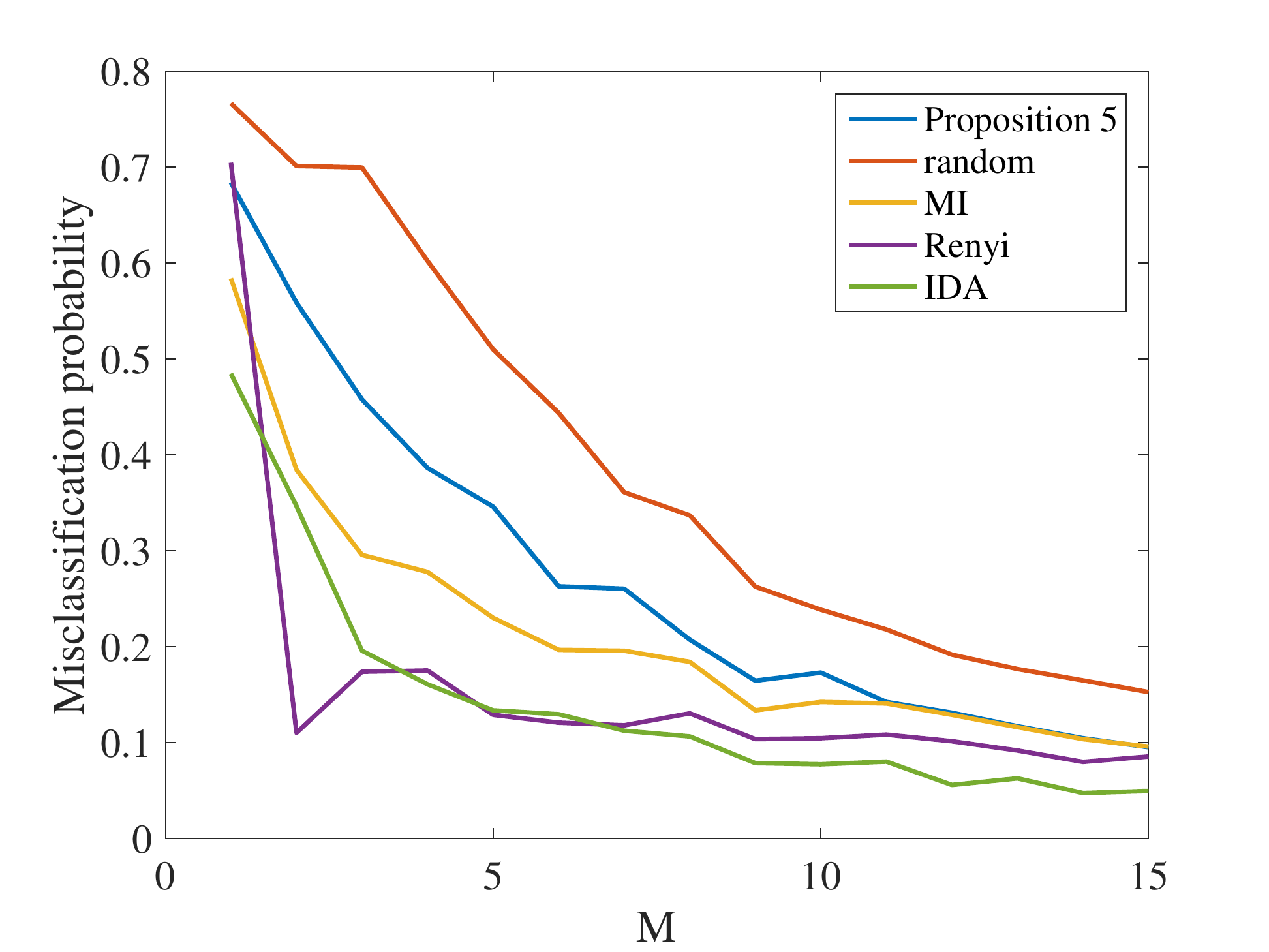}}
\subfigure[$30 \%$ training samples]{\includegraphics[width=0.42\textwidth]{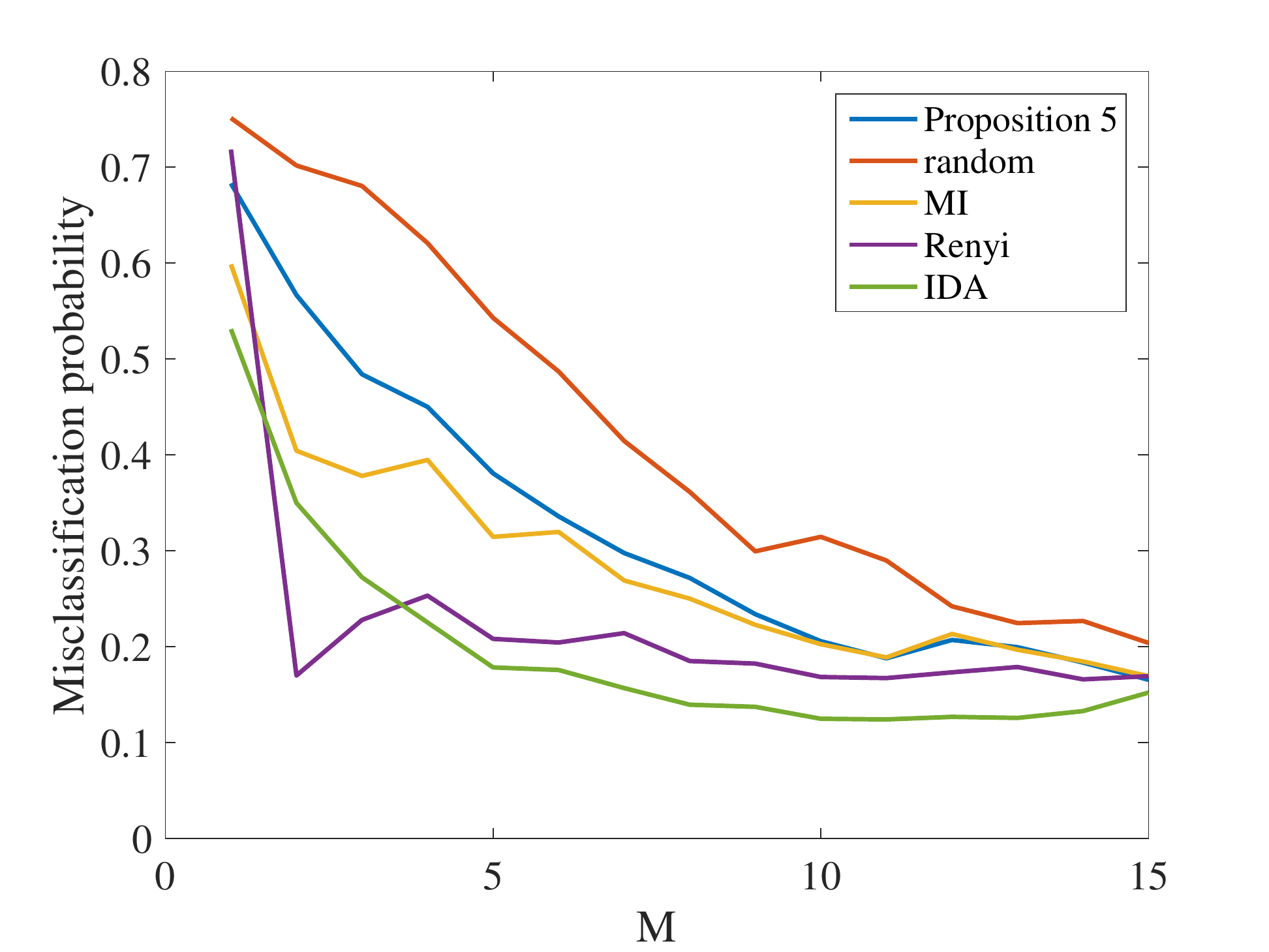}}
%\subfigure[$15 \%$ training samples]{\includegraphics[width=0.42\textwidth]{FacesVsM_015train.eps}}
\end{center}
\caption{Misclassification probability vs. $M$. Extended Yale Face Database B. $50 \%$ or $30 \%$ of the samples are manually labeled and used for training.}
\label{fig:FacesVsM}
\end{figure}

Figures~\ref{fig:FacesVsM} (a) and (b) report the misclassification probability vs the number of measurements for random kernels, kernels designed via the construction embodied in Proposition~\ref{theo:PTmulticlass}, and the designs in \cite{Nenadic07,Chen12}. Also in this case, motivated by the fact that the analysis is conducted for the scenario where the \ac{MAP} classifier is provided with the true model parameters, our results consider both the scenario where a significant number of training samples ($50\%$) is used to learn the underlying models and a scenario where a lower number of training samples ($30\%$) is used to derive the models in order to assess the robustness of the theoretical insights agains model mismatch. We note that in this case, due to the slow eigenvalue decay reported in Fig. \ref{fig:EigFaces}, the measurement design described in Section \ref{par:PTmulticlass} does not provide state-of-the-art classification results, as classification based on measurements extracted via the methods in \cite{Nenadic07,Chen12} guarantee lower misclassification probabilities.

On the other hand, it is possible to observe that the theoretical results in Proposition \ref{prop:1} and Proposition \ref{theo:PTmulticlass} indeed capture the actual behavior of classification with state-of-the-art measurement design. In fact, the upper bounds \eqref{eq:bound_random} \eqref{eq:UB_PT_multiclass} applied to the face recognition scenario under exam predicts that $M=r_{\mathbf{\Sigma}}+1=10$ random measurements or $M=L-1=4$ designed measurements are required for reliable classification. Then, based on numerical simulations of classification with non-compressive measurements, we set the baseline misclassification probability for reliable classification at $25\%$. We observe that the predictions offered by Proposition \ref{prop:1} and Proposition \ref{theo:PTmulticlass} are in line with the trends shown in Table~\ref{table:Faces}, which reports the minimum number of measurements required by  random and designed kernels to achieve a misclassification probability below $25\%$ {for both cases when $50 \%$ and $30\%$ of the vectors in the dataset are used as training samples}.

\begin{table}
\caption{Minimum number of measurements $M$ required to achieve $P_e <25 \%$. Extended Yale Face Database B, $L=5$.}
\begin{center}
\begin{tabular}{c||c|c|c|c|c}
training     & random & Prop. \ref{theo:PTmulticlass} & IDA & MI & R\'enyi \\
  \hline
$50 \%$ &  10 & 8  & 3  & 5 & 2   \\
$30 \%$ &  12 & 9 & 4  & 8 & 2 
\end{tabular}
\end{center}
\label{table:Faces}
\end{table}

}

{\section{Discussion: Impact of Model Mismatch}
\label{par:discussion}

It is also instructive to discuss the impact of model mismatch on the classification performance of the \ac{MAP} classifier \eqref{class} in practical application scenarios. 

In fact, %the classifier \eqref{class} is shown to minimize the average error probability provided that the probabilities $p(C=i|\mathbf{y})$ represent the true distributions from which the observation vectors are drawn \cite{duda00}. Also, 
the analysis carried out in the previous sections assumed that the MAP classifier is given the true model parameters. On the other hand, in practical applications, the conditional \ac{pdf}s $p(\mathbf{y}|C=i)$ and the prior probabilities $p_i$ are usually learnt from training data, thus implying the introduction of mismatch between the model adopted by the classifier and the actual statistical description of test data. 

A proper derivation of the number of measurements required for reliable classification in practical application scenarios would therefore require a more in-depth analysis that takes into account the model mismatch induced by the learning process. In particular, it would require: i) expressions that articulate about the behaviour of the misclassification probability as a function of the true underlying model and the learnt model; ii) a further analysis that determines how compressive random or designed measurements influence the phase transition associated with the misclassification probability.

A preliminary analysis of the impact of model mismatch in classification problems has been conducted in~\cite{Sokolic15,Sokolic16}. These works consider the classification of signals drawn from Gaussian distributions with mismatched classifiers. In particular, they provide sufficient conditions on the relationship between the true model parameters and the learnt model parameters that guarantee reliable classification in the low-noise regime. However, the results in~\cite{Sokolic15,Sokolic16} are derived for a non-compressive classification scenario, therefore they cannot explain how compressive random or designed measurements influence the misclassification probability.

A generalization of our analysis on the minimum number of measurements sufficient for reliable classification to capture the impact of model mismatch does not seem immediate. However, our simulation results associated with real-data subspace classification problems in Sections \ref{par:Hopkins} and \ref{par:Faces} suggest that our theory can still provide meaningful insights both in the situation where we use a significant number of training samples (as expected because we can learn an accurate data model) and in the situation where we use a lower number of training samples. This is despite the fact that the learning process produces distributions that do not correspond exactly to the true ones and also the modelling process assumes a Gaussian distribution that does not necessarily correspond to the true ones pertaining to the motion segmentation or face classification examples.

We conjecture that the reasons for this phenomenon are related to the fact that reliable classification is achieved when compressive measurements are able to discriminate among linear subspaces spanned by signals in the different classes, irrespectively to the particular shape of the marginal distributions that are supported on such subspaces.

In this sense, motion segmentation is more immune to model mismatch than face recognition because, as it is implied by the quick decay of the eigenvalues of the covariance matrices, the majority of the energy of the samples in the motion segmentation dataset is concentrated in linear subspaces of dimension 2 or 3. Then, even a reduced number of training samples is sufficient to identify the dominating principal components for each class. On the other hand, when considering face recognition, the energy associated to samples drawn from a given class is only approximately concentrated on a low-dimensional subspace. In this case, training sets with increased cardinality can guarantee a refined estimation of the principal components associated to each class.

}

\section{Conclusions}
\label{par:conclusions}

In this paper we have offered a characterization of the number of measurements required to reliably classify linear subspaces modeled via low-rank, zero-mean Gaussian distributions. In particular, we have provided upper bounds to the number of measurements required to drive the misclassification probability to zero both for random {measurements} as well as designed {measurements} for two-class classification problems and more challenging multi-class problems. Our characterization suggests that the minimum number of measurements required for phase transition may be achieved by either a \emph{one-vs-all} approach, or by randomly spreading measurements over the Grassmann manifold, depending on the relationship between the number of classes and the dimension of the spaces spanned by signals in each class.

One of the hallmarks of our characterizations relates to its ability to predict the minimum number of measurements required to achieve a low-misclassification probability in state-of-the-art {measurement} design methods. Therefore, it offers engineers a concrete tool to gauge the number of measurements for reliable classification, thereby bypassing the need for time-consuming simulations.

%We also observe that a generalization of the analysis carried out in this paper to the case when signals in each class are described by a \ac{GMM} can be obtained by leveraging the upper bound on the misclassification probability described in \cite{Renna14si}. In this case, conditions for phase transition must hold for all possible pairs of Gaussian distributions associated to classes with distinct labels. \commentFR{not sure I want to keep this last paragraph}

\appendices

\section{Proof of Lemma \ref{theorem1}}
\label{app_A}

Consider the eigenvalue decomposition of the following matrices {$\mathbf{S}_i={\bf \Phi}{\bf \Sigma}_i{\bf \Phi}^T $, $\mathbf{S}_j={\bf \Phi}{\bf \Sigma}_j{\bf \Phi}^T$ and $\mathbf{S}_{ij}={\bf \Phi}\left({\bf \Sigma}_i+{\bf \Sigma}_j\right){\bf \Phi}^T $, which yields
\begin{IEEEeqnarray}{rCl}
\mathbf{S}_i & = & {\bf U}_i{\bf \Lambda}_i {\bf U}_i^{T} \\
\mathbf{S}_j & = & {\bf U}_j{\bf \Lambda}_j {\bf U}_j^{T} \\
\mathbf{S}_{ij} & = & {\bf U}_{ij}{\bf \Lambda}_{ij} {\bf U}_{ij}^{T}.
\end{IEEEeqnarray}
 where $\mathbf{U}_i, \mathbf{U}_j,\mathbf{U}_{ij} \in \mathbb{R}^{M \times M}$ are orthogonal matrices; the diagonal matrices ${\bf \Lambda}_i = \mathrm{diag}(\lambda_{i_1},\cdots,\lambda_{i_{r_i}},0,\cdots,0)$, ${\bf \Lambda}_j = \mathrm{diag}(\lambda_{j_1},\cdots,\lambda_{j_{r_j}},0,\cdots,0)$ and  ${\bf \Lambda}_{ij} = \mathrm{diag}(\lambda_{{ij}_1},\cdots,\lambda_{{ij}_{r_{ij}}},0,\cdots,0)$ contain the eigenvalues of $\mathbf{S}_i, \mathbf{S}_j$ and $\mathbf{S}_{ij}$, respectively. Note that the number of strictly positive eigenvalues of $\mathbf{S}_i, \mathbf{S}_j$ and $\mathbf{S}_{ij}$, i.e., the number of strictly positive diagonal entries in $\mathbf{\Lambda}_i, \mathbf{\Lambda}_j$ and $\mathbf{\Lambda}_{ij}$, is equal to $r_i = \mathrm{rank}(\mathbf{S}_i)=\mathrm{rank}(\mathbf{\Phi} \mathbf{\Sigma}_i \mathbf{\Phi}^T), r_j= \mathrm{rank}(\mathbf{S}_j)=\mathrm{rank}(\mathbf{\Phi} \mathbf{\Sigma}_j\mathbf{\Phi}^T)$ and $r_{ij}= \mathrm{rank}(\mathbf{S}_{ij})=\mathrm{rank}(\mathbf{\Phi} (\mathbf{\Sigma}_i  + \mathbf{\Sigma}_j )\mathbf{\Phi}^T)$, respectively. 
 
% with $r_i = \mathrm{rank} \left({\bf \Phi}{\bf \Sigma}_i{\bf \Phi}^T\right)$, and $r_{ij} = \mathrm{rank} \left({\bf \Phi}\left({\bf \Sigma}_i+{\bf \Sigma}_j\right){\bf \Phi}^T\right)$.
%
Then, we recall the expression of the upper bound to the misclassification probability
\begin{equation}
\bar{P}_e = \sum_{i=1}^L  \sum_{\substack{j=1 \\ j\neq i}}^L \sqrt{p_i p_j} e^{-K_{ij}},
\end{equation}
and we can re-express $K_{ij}$ as
\begin{IEEEeqnarray}{rCl}
\label{app_a_2} 
K_{ij} & = & \frac{1}{4}\log\frac{\left(   \mathrm{det} \left(\frac{{\bf \Phi}\left({\bf \Sigma}_i+{\bf \Sigma}_j\right){\bf \Phi}^T + 2 \sigma^2 \mathbf{I}}{2}\right)  \right)^2}{\mathrm{det} \left({\bf \Phi}{\bf \Sigma}_i{\bf \Phi}^T + \sigma^2 \mathbf{I}\right)  \mathrm{det} \left({\bf \Phi}{\bf \Sigma}_j{\bf \Phi}^T + \sigma^2 \mathbf{I}\right)} \nonumber \\ 
& = & \frac{1}{4}\log\frac{\left(   \mathrm{det} \left(\frac{\mathbf{S}_{ij} + 2 \sigma^2 \mathbf{I}}{2}\right)  \right)^2}{\mathrm{det} \left(\mathbf{S}_i + \sigma^2 \mathbf{I}\right)  \mathrm{det} \left(\mathbf{S}_j + \sigma^2 \mathbf{I}\right)} \nonumber \\ 
\nonumber
& = & \frac{1}{4}\log \left[2^{-2r_{ij}} {\left(\sigma^2\right)}^{r_i + r_j - 2r_{ij}}  \right. \\
&& \left.  \cdot \frac{   \prod_{k=1}^{r_{ij}}\left(\lambda_{{ij}_k} + 2\sigma^2\right)^2    }{  \prod_{k=1}^{r_{i}}\left(\lambda_{{i}_k} + \sigma^2\right)       \prod_{k=1}^{r_{j}}\left(\lambda_{{j}_k} + \sigma^2\right)    }\right],
\label{eq:Kexpanded}
\end{IEEEeqnarray}
thus leading to 
\begin{IEEEeqnarray}{rCl}
\nonumber
\bar{P}_e & = & \sum_{i=1}^L  \sum_{\substack{j=1 \\ j\neq i}}^L \sqrt{p_i p_j} \, 2^{r_{ij}/2} (\sigma^2)^{(2r_{ij}-r_i -r_j)/4}  \\
& &\cdot \left[  \frac{ \sqrt{  \prod_{k=1}^{r_{i}}\left(\lambda_{{i}_k} + \sigma^2\right)       \prod_{k=1}^{r_{j}}\left(\lambda_{{j}_k} + \sigma^2\right)      }   }{\prod_{k=1}^{r_{ij}}\left(\lambda_{{ij}_k} + 2\sigma^2\right)^2 }  \right]^{1/2}.
\end{IEEEeqnarray}
Then, on letting $\sigma^2 \to 0$, we note that the term in square brackets converges to the positive constant $\left[  \frac{\sqrt{v_i v_j}}{v_{ij}} \right]^{1/2}$, moreover, the decay of $\bar{P}_e$ as a function of $\sigma^2$ is dominated by the terms in the sum corresponding to the minimum value of the exponent $d(i,j)=(2r_{ij}-r_i -r_j)/4$, thus leading to the result in \eqref{eq:expansion}-\eqref{gm_mult}.
%
%Then, the asymptotic expansion of the upper bound to the misclassification probability follows immediately by substituting \eqref{eq:Kexpanded} in  \eqref{multiclass} and by letting $\sigma^2 \to 0$.
}

\section{Proof of Proposition \ref{theorem4}}
\label{app_D}

The derivation of the upper bound on the number of measurements needed to verify (\ref{eq:d_prop4}) is based on the analysis of the upper bound $\bar{P}_{e}$ in (\ref{multiclass}).

Recall that the low-noise expansion exponent $d$ of the upper bound to the misclassification probability for the classification problem of two, zero-mean classes is given by $d =  \left(2r_{12} - r_1 - r_2\right)/4$.
%\begin{equation}
%\label{diver_design}
%d =  \left(2r_{12} - r_1 - r_2\right)/4 .
%\end{equation}

We first show that, for all possible choices of $\mathbf{\Phi}$, it holds $d\leq R/4$ so that there is not any $M$ such that
\begin{equation}
\lim_{\sigma^2 \to 0}  -\frac{\log \bar{P}_e}{\log(1/\sigma^2)} >d_0
\label{eq:d_upperbound}
\end{equation}
for $d_0\geq R/4$. 

Then, we consider the case $d_0<R/4$ and we derive the minimum number of measurements $M$ needed to verify \eqref{eq:d_upperbound}, which represents an upper bound on the minimum number of measurements needed to verify (\ref{eq:d_prop4}).

%We first show that the feasible set of the problem \eqref{eq:PT2classes} is empty when $d_0 \geq R/4$ and then we characterize the solution of the problem for the case $d_0 < R /4$.

\subsection{Case where $d_0 \geq R/4$}
\label{app:upper}

Let $r_{\mathbf{\Sigma}_{12}} = \mathrm{rank}(\mathbf{\Sigma}_1 + \mathbf{\Sigma}_2)$. In the following, we show that, for all possible choices of $\mathbf{\Phi}$, it holds
\begin{equation}
d=(2r_{12} - r_1 - r_2)/4 \leq (2r_{\B{\Sigma}_{12}} - 2 r_{\B{\Sigma}} )/4 =  R/4,
\label{r_rsigma}
\end{equation}
or, instead,
\begin{equation}
r_{\B{\Sigma}_{12}} - r_{12} \geq r_{\B{\Sigma}} - r_1 \wedge r_{\B{\Sigma}_{12}} - r_{12} \geq r_{\B{\Sigma}}  - r_2,
\label{nes_suf_cond}
\end{equation}
since (\ref{nes_suf_cond}) implies (\ref{r_rsigma}). Consider the generalized eigenvalue decomposition of the positive semidefinite matrices $\mathbf{\Sigma}_1$ and $\mathbf{\Sigma}_2$ given by~\cite[Theorem 8.7.1]{Golub96}, namely,
\begin{equation}  
{\B{\Sigma}_{1}} = \mathbf{X}^{-T}\mathbf{D}_1\mathbf{X}^{-1} = \mathbf{X}^{-T}\ \mathrm{diag}\left(d_{1_1}, \ldots, d_{1_N}\right)\mathbf{X}^{-1},
\end{equation} 
with $d_{1_i} \geq 0, i = 1, \ldots, N$ and 
\begin{equation}
{\B{\Sigma}_{2}} = \mathbf{X}^{-T}\mathbf{D}_2\mathbf{X}^{-1} = \mathbf{X}^{-T}\ \mathrm{diag}\left(d_{2_1}, \ldots, d_{2_N}\right)\mathbf{X}^{-1},
\end{equation}
 with $d_{2_i} \geq 0, i = 1, \ldots, N$, where $\mathbf{X}$ is a non-singular matrix. 

Note that we have
\begin{IEEEeqnarray}{rCl}
r_{12} & = & \mathrm{rank} \left( \B{\Phi}\mathbf{X}^{-T}\left(\mathbf{D}_1 + \mathbf{D}_2\right)\mathbf{X}^{-1}\B{\Phi}^{-T}\right)\\
& =&  \mathrm{rank} \left( \B{\tilde{\Phi}}\left(\mathbf{D}_1 + \mathbf{D}_2\right)^{\frac{1}{2}}\right)
\end{IEEEeqnarray}
 and likewise, 
 \begin{IEEEeqnarray}{rCl}
 r_{1} & = & \mathrm{rank} \left( \B{\Phi}\mathbf{X}^{-T}\mathbf{D}_1 \mathbf{X}^{-1}\B{\Phi}^{-T}\right)
=  \mathrm{rank} \left( \B{\tilde{\Phi}}\mathbf{D}_1^{\frac{1}{2}}\right) \IEEEeqnarraynumspace
 \end{IEEEeqnarray}
 and 
 \begin{IEEEeqnarray}{rCl}
 r_{2} & = & \mathrm{rank} \left( \B{\Phi}\mathbf{X}^{-T}\mathbf{D}_2 \mathbf{X}^{-1}\B{\Phi}^{-T}\right) =\mathrm{rank} \left( \B{\tilde{\Phi}}\mathbf{D}_2^{\frac{1}{2}}\right),  \IEEEeqnarraynumspace
 \end{IEEEeqnarray}
where $\B{\tilde{\Phi}} = \B{\Phi}\mathbf{X}^{-T}$.

On the other hand, the ranks of the input covariance matrices can be expressed as 
\begin{IEEEeqnarray}{rCl}
r_{\B{\Sigma}_{12}} & = & \mathrm{rank} \left(\mathbf{X}^{-T}\left(\mathbf{D}_1 + \mathbf{D}_2\right)\mathbf{X}^{-1}\right)\\
& =& \mathrm{rank} \left( \mathbf{X}^{-T}\left(\mathbf{D}_1 + \mathbf{D}_2\right)^{\frac{1}{2}}\right)\\
& = & \mathrm{rank} \left( \left(\mathbf{D}_1 + \mathbf{D}_2\right)^{\frac{1}{2}}\right) 
\end{IEEEeqnarray}
 and
 \begin{IEEEeqnarray}{rCl}
 r_{\B{\Sigma}} & = & \mathrm{rank} \left( \left(\mathbf{D}_1 \right)^{\frac{1}{2}}\right) =  \mathrm{rank} \left( \left(\mathbf{D}_2 \right)^{\frac{1}{2}}\right).
\end{IEEEeqnarray}

Let us now define the cardinalities of the following sets: 
\begin{IEEEeqnarray}{rCl}
k_c &= & \left|\left\{i:d_{1_i} > 0 \wedge d_{2_i} > 0\right\}\right|\\
k_1 & =&  \left|\left\{i: d_{1_i} > 0 \right\}\right|\\
k_2 & = &  \left|\left\{i: d_{2_i} > 0\right\}\right|.
\end{IEEEeqnarray} 
Then, it becomes evident that, $r_{\B{\Sigma}_{12}} -  r_{\B{\Sigma}} = k_1 + k_2 - k_c - k_1 = k_2 - k_c= k_1 - k_c$, and, in view of the possible dependence between columns of $\B{\tilde{\Phi}}$, $r_{12} -  r_{1} \leq k_2 - k_c$ and $r_{12} -  r_{2} \leq k_1 - k_c$, thus concluding the proof of (\ref{r_rsigma}).

\subsection{Case where $d_0 < R/4$}

We start by describing an explicit {measurement} matrix construction that achieves an expansion exponent of the upper bound to the misclassification probability strictly greater than $d_0$ with $M=\lfloor 4 d_0 \rfloor +1$ measurements. After that, we prove that $M\leq \lfloor 4 d_0 \rfloor $ implies $d \leq d_0$ for all possible choices of $\mathbf{\Phi}$.

\subsubsection{Achievability}
%This upper bound -- in view of (\ref{r_rsigma}) -- can be achieved by a projections matrix design that satisfies
%\begin{equation}
%\label{cond_desig_zero}
%R = 2r_{12} - r_1 - r_2  = 2r_{\B{\Sigma}_{12}} - r_{\B{\Sigma}_{1}} - r_{\B{\Sigma}_{2}} = {n_{\B{\Sigma}_{1}}} + {n_{\B{\Sigma}_{2}}},
%\end{equation}
%where we have used the fact that $r_{\B{\Sigma}_{1}} = N - \mathrm{dim}\left(\mathrm{Null}\left({\B{\Sigma}_{1}}\right)\right)= N - n_{12}-{n_{\B{\Sigma}_{1}}}$, $r_{\B{\Sigma}_{2}} = N - \mathrm{dim}\left(\mathrm{Null}\left({\B{\Sigma}_{2}}\right)\right)= N - n_{12}-{n_{\B{\Sigma}_{2}}}$ and $r_{\B{\Sigma}_{12}} = N - \mathrm{dim}\left(\mathrm{Null}\left(\B{\Sigma}_{1}\right)  \bigcap \mathrm{Null}\left(\B{\Sigma}_{2}\right)\right)= N - n_{12}$, where $n_{12}$, $n_{12}+n_{\B{\Sigma}_{1}}$ and  $n_{12}+n_{\B{\Sigma}_{2}}$ are the dimensions of the sub-spaces $\mathrm{Null}\left(\B{\Sigma}_{1}\right)  \bigcap \mathrm{Null}\left(\B{\Sigma}_{2}\right)$, $\mathrm{Null}\left(\B{\Sigma}_{1}\right)$ and $\mathrm{Null}\left(\B{\Sigma}_{2}\right)$, respectively. Note that, in order to guarantee \eqref{cond_desig_zero}, the two conditions in \eqref{nes_suf_cond} have to hold with equality, thus implying that $r_2 \geq 2r_{12} - r_1 = r_{\B{\Sigma}_{12}} - r_{\B{\Sigma}_{1}}$ and $r_1 \geq 2r_{12} - r_2 = r_{\B{\Sigma}_{12}} - r_{\B{\Sigma}_{2}}$.

Consider the matrix
\begin{equation}
\B{\Phi}_0 = \left[\B{v}_1, \B{v}_2,	\ldots, \B{v}_{n_{\B{\Sigma}}},\B{w}_1,	\B{w}_2,\ldots,\B{w}_{n_{\B{\Sigma}}}\right]^T,
\label{phi_opt}
\end{equation}
where the sets $\left[\B{u}_1,\ldots,\B{u}_{n_{12}}\right]$, $\left[\B{u}_1,\ldots,\B{u}_{n_{12}},\B{v}_1,\ldots,\B{v}_{n_{\B{\Sigma}}}\right]$, $\left[\B{u}_1,\ldots,\B{u}_{n_{12}},\B{w}_1,\ldots,\B{w}_{n_{\B{\Sigma}}}\right]$,  $\B{u}_i, \B{v}_i, \B{w}_i  \in \mathbb{R}^N$, constitute an orthonormal basis of the linear spaces $\mathcal{N}_{12} = \mathrm{Null}\left(\B{\Sigma}_{1}\right)  \bigcap \mathrm{Null}\left(\B{\Sigma}_{2}\right)$, $\mathcal{N}_1 =\mathrm{Null}\left(\B{\Sigma}_{1}\right)$ and $\mathcal{N}_2=\mathrm{Null}\left(\B{\Sigma}_{2}\right)$, respectively, and $n_{12}=[N-2r_{\mathbf{\Sigma}}]^+$, $n_{\B{\Sigma}} =  \min \{  N-r_{\mathbf{\Sigma}}, r_{\mathbf{\Sigma}}  \}=R/2$.

Then, we can write
\begin{equation}
\mathbf{\Phi}_0 \mathbf{\Sigma}_1 \mathbf{\Phi}_0^T=\left[
\begin{array}{c|c}
  \B{0} & \B{0} \\
  \hline
  \B{0} & \mathbf{Q}
 \end{array}\right] 
%\mathbf{Q} = \begin{bmatrix}
%	\B{w}_1   & \B{w}_2	& \cdots & \B{w}_{n_{\B{\Sigma}_{2}}} \end{bmatrix}^T  \B{\Sigma}_{1} \begin{bmatrix}
%	\B{w}_1   & \B{w}_2	& \cdots & \B{w}_{n_{\B{\Sigma}_{2}}} \end{bmatrix}
\end{equation}
where
\begin{equation}
\mathbf{Q} = [\mathbf{w}_1,\ldots,\mathbf{w}_{n_{\mathbf{\Sigma}}}]^T \B{\Sigma}_1[\mathbf{w}_1,\ldots,\mathbf{w}_{n_{\mathbf{\Sigma}}}],
\end{equation}
so that we also have $r_1=\mathrm{rank}\left(\mathbf{\Phi}_0 \mathbf{\Sigma}_1 \mathbf{\Phi}_0^T\right) = \mathrm{rank}\left(\mathbf{Q} \right)$. Now, note that the matrix $\mathbf{Q}$ is the Gram matrix of the set of vectors $\B{q}_i= \B{\Sigma}_{1}^{\frac{1}{2}}\B{w}_i$, $i=1, \ldots, {n_{\B{\Sigma}}}$, and, therefore, $r_1=\mathrm{rank}\left(\mathbf{Q} \right) = {n_{\B{\Sigma}}}$ if and only if the vectors $\B{q}_i$, $i=1, \ldots, {n_{\B{\Sigma}}}$, are linearly independent.

Assume by contradiction that the vectors $\B{q}_i$ are linearly dependent. Then, there exists a set of $n_{\mathbf{\Sigma}}$ scalars $\alpha_i$ (with $\alpha_i \neq 0$ for at least one index $i$) such that $\B{\Sigma}_{1}^{\frac{1}{2}} \sum_i \alpha_i \B{w}_i = \mathbf{0}$. It is known that $\sum_i \alpha_i \B{w}_i \neq \mathbf{0}$ because $\B{w}_i$ are linearly independent by construction. Therefore, the linearly dependence among the vectors $\B{q}_i$ implies that $\sum_i \alpha_i \B{w}_i \in \mathcal{N}_1$, 
%\begin{equation}
%\sum_i \alpha_i \B{w}_i \in \mathrm{Null}\left(\B{\Sigma}_{1}^{\frac{1}{2}}\right)\  \mbox{or}  \  \sum_i \alpha_i \B{w}_i \in \mathrm{Null}\left(\B{\Sigma}_{1}\right)
%\end{equation}
which is false since, by construction, $\sum_i \alpha_i \B{w}_i \in \mathcal{N}_2$ and $\sum_i \alpha_i \B{w}_i \notin \mathcal{N}_{12}$. Therefore, we can establish that $r_1 =\mathrm{rank}\left(\mathbf{\Phi}_0 \mathbf{\Sigma}_1 \mathbf{\Phi}_0^T\right) = \mathrm{rank}\left(\mathbf{Q} \right) = {n_{\B{\Sigma}}}$, and,  we can similarly establish that $r_2 =\mathrm{rank}\left(\mathbf{\Phi}_0 \mathbf{\Sigma}_2 \mathbf{\Phi}_0^T\right) =  {n_{\B{\Sigma}}}$ and $r_{12} = \mathrm{rank}\left(\mathbf{\Phi}_0 (\mathbf{\Sigma}_i + \mathbf{\Sigma}_j) \mathbf{\Phi}_0^T\right) = 2{n_{\B{\Sigma}}} =R$.
%, that is, this matrix construction, which satisfies the condition in (\ref{cond_desig_zero}), achieves the maximum diversity-order in (\ref{max_div_ord}).

Finally, we generate $\mathbf{\Phi}$ by picking arbitrarily only $M=\lfloor 4 d_0 \rfloor +1$ among the $R$ row vectors of the matrix $\mathbf{\Phi}_0$ in (\ref{phi_opt}). In particular, we take $M_1$ rows from the set $\left[ \mathbf{v}_1,\ldots, \mathbf{v}_{n_{\mathbf{\Sigma}}}\right]$ and $M_2$ rows from the set $\left[ \mathbf{w}_1,\ldots, \mathbf{w}_{n_{\mathbf{\Sigma}}}\right]$, where $M_1 + M_2=\lfloor 4 d_0 \rfloor +1$, which is always possible as $\lfloor 4 d_0 \rfloor +1 \leq R$. Then, by following steps similar to the previous ones, it is possible to show that $r_1 = \mathrm{rank}\left(\mathbf{\Phi} \mathbf{\Sigma}_1 \mathbf{\Phi}^T\right) =  M_2$, $r_2 = \mathrm{rank}\left(\mathbf{\Phi} \mathbf{\Sigma}_2 \mathbf{\Phi}^T\right) =  M_1$ and $r_{12} = \mathrm{rank}\left(\mathbf{\Phi} (\mathbf{\Sigma}_1 + \mathbf{\Sigma}_2) \mathbf{\Phi}^T\right) = M_1+M_2$, thus implying $d=(\lfloor 4 d_0 \rfloor +1)/4>d_0$.

\subsubsection{Converse}
Assume now $M \leq \lfloor 4 d_0  \rfloor$. In this case, we can show that, for all possible choices of $\mathbf{\Phi}$, it holds
\begin{equation}
d \leq M/4 \leq d_0.
\end{equation}
This upper bound follows from the solution to the following integer-valued optimization problem\footnote{Note that this problem represents a relaxation of the problem which aims at maximizing $d$, as it incorporates only some of the constraints dictated by the geometrical description of the scenario. For example, it does not take into account the actual value of some parameters of the input description as $r_{\mathbf{\Sigma}}$ and $r_{\mathbf{\Sigma}_{12}}$.
}:
\begin{equation}
\label{opt_1}
\max_{r_{1}, r_{2}, r_{12}}  \left( 2r_{12} -{r_{1}-r_{2}} \right)/4
\end{equation}
subject to: $r_{1}+r_{2} \geq r_{12}$, $r_{1} \leq M$, $r_{2} \leq M$, $r_{12} \leq M$ and $r_{1}, r_{2}, r_{12} \in \mathbb{Z}_0^+$.

The solution, which can be obtained by considering a linear programming relaxation along with a Branch and Bound approach~\cite{Schrijver98}, is given by\footnote{The solution of the optimization problem is not unique. Nevertheless, the maximum value achieved by the objective function is indeed unique.}:
\begin{equation}
r_{1}+ r_{2}=r_{12}
\qv
r_{12} = M,
\end{equation}
\begin{equation}
d =\left( 2r_{12}-{r_{1}-r_{2}}\right)/4 =  M/4.
\label{up_div}
\end{equation}

\section{Proof of Proposition \ref{theo:PTmulticlass}}
\label{app:design_multiclass}

Let $\mathbf{N}_i \in \mathbb{R}^{N \times (N-r_{\B{\Sigma}})}$ be a matrix that contains a basis for the null space $\mathcal{N}_i$ and let $\mathbf{N} = [\mathbf{N}_1, \ldots, \mathbf{N}_L]$ be a matrix that contains the concatenation of the bases for all the null spaces $\mathcal{N}_1,\ldots,\mathcal{N}_L$. Then, consider the {measurement} matrices $\mathbf{\Phi} \in \mathbb{R}^{M \times N}$ that consist of $M$ rows of $\mathbf{N}^T$. More precisely, such matrices  $\mathbf{\Phi}$ are obtained by picking $M_i$ rows from $\mathbf{N}_i^T$, so that $\sum_{i=1}^L M_i =M$.\footnote{Throughout the proof, we assume $M\leq N$, since the decay exponent $d$ associated to any matrix $\mathbf{\Phi}$ is always smaller than or equal to the decay exponent associated to the identity matrix $\mathbf{I}_N$, as it was shown in Appendix~\ref{app:upper}.}

%In the following, we will show that it is possible to re-express the constraint in problem \eqref{eq:PT2classes} as a set of constraints on the values $M_i$. 

A sufficient condition for \eqref{eq:PT_multiple} is represented by $d>0$, where $d$ is the decay exponent associated to the misclassification probability upper bound \eqref{multiclass}. Moreover, $d>0$ if and only if $d(i,j)>0$ for all the pairs $(i,j)$ with $i \neq j$. 

We can now express the conditions $d(i,j)>0$ in terms of the values $M_i$ as follows. 
On recalling Sylvester's rank theorem \cite{Meyer00}, which states
\begin{equation}
\rank{ \mathbf{A} \mathbf{B}} = \rank{\mathbf{B}} - \dim(\mathrm{Im}(\mathbf{B})   \cap \mathrm{Null}(\mathbf{A})),
\end{equation}
we can write each term $d(i,j)$ as
\begin{IEEEeqnarray}{rCl}
d(i,j) & = & \left(2 r_{ij} - r_i - r_j\right)/4 \\
\nonumber
 & = & \left[ \dim (\mathrm{Im} (\mathbf{\Phi}^T)\cap \mathcal{N}_i) + \dim (\mathrm{Im} (\mathbf{\Phi}^T)\cap \mathcal{N}_j) \right.  \\
 & &   \left. - 2 \dim (\mathrm{Im} (\mathbf{\Phi}^T)\cap \mathcal{N}_{ij})  \right]/4.
 \label{eq:dij_dim}
\end{IEEEeqnarray}
We first show that
\begin{equation}
\dim (\mathrm{Im} (\mathbf{\Phi}^T)\cap \mathcal{N}_i) = \max \{ M - r_{\B{\Sigma}} , M_i  \}.
\label{eq:dim_int_1}
\end{equation}
Notice that, since the images $\mathcal{R}_i$ are independently drawn from a continuous \ac{pdf} over the Grassmann manifold, any $\min \{ N , L(N - r_{\B{\Sigma}})   \} $ columns of $\mathbf{N}$ are linearly independent with probability 1. 
Then, by leveraging the expression of the dimension of the intersection of two linear spaces, we can write
\begin{IEEEeqnarray}{rCl}
\nonumber
\dim (\mathrm{Im} (\mathbf{\Phi}^T)\cap \mathcal{N}_i)& = & \rank{\B{\Phi}} + \rank{\mathbf{N}_i} - \mathrm{rank}[\B{\Phi}^T \, \mathbf{N}_i]\\
& = & M +(N-r_{\B{\Sigma}}) - \mathrm{rank}[\B{\Phi}^T \, \mathbf{N}_i] . \IEEEeqnarraynumspace
\label{eq:dim_int_1_a}
\end{IEEEeqnarray}
%where the columns of the matrix $\mathbf{N}_i \in \mathbb{R}^{N \times (N-r_{\B{\Sigma}})}$ form a basis for $\mathcal{N}_i$. 
Moreover, 
\begin{equation}
\mathrm{rank}[\B{\Phi}^T \, \mathbf{N}_i] = \mathrm{rank}[\bar{\B{\Phi}}^T \, \mathbf{N}_i],
\end{equation}
where $\bar{\B{\Phi}}^T$ is obtained from ${\B{\Phi}}^T$ by deleting the $M_i$ columns corresponding to vectors taken from the basis of the null space $\mathcal{N}_i$. Then, given that the columns of $\bar{\B{\Phi}}^T$ are picked from spaces drawn at random from the Grassmann manifold, we can conclude that 
\begin{equation}
 \mathrm{rank}[\bar{\B{\Phi}}^T \, \mathbf{N}_i] = \min \{ N , M - M_i + N - r_{\B{\Sigma}}   \},
 \label{eq:rank_bar_phi_i}
\end{equation}
and on replacing \eqref{eq:rank_bar_phi_i} into \eqref{eq:dim_int_1_a} we immediately obtain \eqref{eq:dim_int_1}. 

Consider now the last term in \eqref{eq:dij_dim} 
%\begin{equation}
%\dim (\mathrm{Im} (\mathbf{\Phi}^T)\cap \mathcal{N}_{ij}),
%\end{equation}
and recall that, since the linear spaces $\mathcal{N}_i$ are drawn independently at random from a continuous \ac{pdf}, then
\begin{equation}
\dim (\mathcal{N}_{ij}) =   \dim ( \mathcal{N}_i \cap \mathcal{N}_j) = \max\{  N-2 r_{\B{\Sigma}}, 0 \},
\end{equation}
thus implying immediately that $\dim (\mathrm{Im} (\mathbf{\Phi}^T)\cap \mathcal{N}_{ij}) = 0$ if $N \leq 2 r_{\B{\Sigma}}$. Therefore, we assume $N > 2 r_{\B{\Sigma}}$ and we show that 
\begin{equation}
\dim (\mathrm{Im} (\mathbf{\Phi}^T)\cap \mathcal{N}_{ij}) = \max \{   M - 2 r_{\B{\Sigma}} , M_i - r_{\B{\Sigma}}, M_j - r_{\B{\Sigma}}, 0  \}.
\label{eq:dim_int_12}
\end{equation}
In order to do that, we first note that we can leverage the expression of the dimension of the intersection of two linear subspaces to write
\begin{equation}
\dim (\mathrm{Im} (\mathbf{\Phi}^T)\cap \mathcal{N}_{ij})  =  M + (N-2r_{\B \Sigma}) - \mathrm{rank}[\B{\Phi}^T  \mathbf{N}_{ij}],
\end{equation}
where the columns of $ \mathbf{N}_{ij}$ form a basis of the linear space $\mathcal{N}_{ij}$. Let us also write $\mathbf{\Phi}$ as 
\begin{equation}
\mathbf{\Phi} = [ \tilde{\mathbf{\Phi}}^T  \, \mathbf{\Phi}_i^T  \, \mathbf{\Phi}_j^T]^T,
\end{equation}
where the $M_i$ columns of $\mathbf{\Phi}_i^T$ are vectors picked from a basis of $\mathcal{N}_i$, the $M_j$ columns of $\mathbf{\Phi}_j^T$ are vectors picked from a basis of $\mathcal{N}_j$ and the $M-M_i -M_j$ columns of $\tilde{\mathbf{\Phi}}^T$ are vectors picked from the bases of the remaining null spaces. Then, on leveraging again the assumption that the linear spaces associated to the different classes are picked independently at random from a continuous distribution, we can write 
\begin{equation}
\mathrm{rank}[\B{\Phi}^T  \mathbf{N}_{ij}]  = \min\{ N, M-M_i - M_j  + \mathrm{rank}[ \mathbf{\Phi}_i^T  \, \mathbf{\Phi}_j^T  \, \mathbf{N}_{ij} ]     \}.
\end{equation}
On the other hand, on introducing the notation $r_{\mathbf{\Phi}_{ij} \mathbf{N}_{ij}}=\mathrm{rank}[ \mathbf{\Phi}_i^T  \, \mathbf{\Phi}_j^T  \, \mathbf{N}_{ij} ]$ we also have
\begin{IEEEeqnarray}{rCl}
\nonumber
r_{\mathbf{\Phi}_{ij} \mathbf{N}_{ij}}  &= &\mathrm{rank}[ \mathbf{\Phi}_i^T\, \mathbf{N}_{ij}  \, \mathbf{\Phi}_j^T  \, \mathbf{N}_{ij} ]     \\
\nonumber
 & = & \dim(\mathrm{Im} [\mathbf{\Phi}_i^T \mathbf{N}_{ij}]) + \dim(\mathrm{Im} [\mathbf{\Phi}_j^T \mathbf{N}_{ij}]) \\
\nonumber
 && - \dim (\mathrm{Im} [\mathbf{\Phi}_i^T \mathbf{N}_{ij}]  \cap \mathrm{Im} [\mathbf{\Phi}_j^T \mathbf{N}_{ij}]) \\
 \nonumber
 &=& \min\{ N - r_{\B{\Sigma}}, M_i + N - 2 r_{\B{\Sigma}}   \}\\
\nonumber
 && + \min\{ N - r_{\B{\Sigma}}, M_j + N - 2 r_{\B{\Sigma}}   \}  - (N-2r_{\B{\Sigma}}). \IEEEeqnarraynumspace
\end{IEEEeqnarray}
In fact, $\dim(\mathrm{Im} [\mathbf{\Phi}_i^T\, \mathbf{N}_{ij}]) = \min\{ N - r_{\B{\Sigma}}, M_i + N - 2 r_{\B{\Sigma}}   \}$ derives from the fact that the columns of $\mathbf{\Phi}_i^T$ and $\mathbf{N}_{ij}$ are all picked at random from the space $\mathcal{N}_i$ -- which has dimension $N-r_{\B{\Sigma}}$. Moreover, we have used the fact
\begin{equation}
 \dim (\mathrm{Im} [\mathbf{\Phi}_i^T \mathbf{N}_{ij}]  \cap \mathrm{Im} [\mathbf{\Phi}_j^T \mathbf{N}_{ij}])  = N-2r_{\B{\Sigma}},
\end{equation}
which follows from 
\begin{equation}
\mathrm{Im} (\mathbf{N}_{ij})  \subseteq\mathrm{Im} [\mathbf{\Phi}_i^T \mathbf{N}_{ij}]  \cap \mathrm{Im} [\mathbf{\Phi}_j^T \mathbf{N}_{ij}]  \subseteq \mathcal{N}_i \cap \mathcal{N}_j.
\end{equation}
Then, on using the symbol $r_{\mathbf{\Phi} \mathbf{N}_{ij}}=\mathrm{rank}[\B{\Phi}^T  \mathbf{N}_{ij}] $, we have
\begin{IEEEeqnarray}{rCl}
\nonumber
r_{\mathbf{\Phi} \mathbf{N}_{ij}} & = &  \min\{ N, M-M_i - M_j \\
\nonumber
&& +  \min\{ N - r_{\B{\Sigma}}, M_i + N - 2 r_{\B{\Sigma}}  \} \\
\nonumber
&& + \min\{ N - r_{\B{\Sigma}}, M_j + N - 2 r_{\B{\Sigma}}   \}   - (N-2r_{\B{\Sigma}})\} \IEEEeqnarraynumspace
\end{IEEEeqnarray}
and, therefore,
\begin{IEEEeqnarray}{rCl}
\nonumber
\dim (\mathrm{Im} (\mathbf{\Phi}^T)\cap \mathcal{N}_{ij}) & = & \max\{   M - 2 r_{\B\Sigma},  M_i + M_j   \\
\nonumber
&& +   \min\{ N - r_{\B{\Sigma}}, M_i + N - 2 r_{\B{\Sigma}}  \}\\
\nonumber
&& + \min\{ N - r_{\B{\Sigma}}, M_j + N - 2 r_{\B{\Sigma}}   \} \\
 &&  - (N-2r_{\B{\Sigma}})\}.
 \label{eq:dim_int_12a}
\end{IEEEeqnarray}
Finally, it is possible to show that \eqref{eq:dim_int_12a} is equivalent to \eqref{eq:dim_int_12} by considering separately the cases for which $M_i \lesseqgtr r_{\B{\Sigma}}$ and $M_j \lesseqgtr r_{\B{\Sigma}}$.

Therefore, by using \eqref{eq:dij_dim}, \eqref{eq:dim_int_1} and \eqref{eq:dim_int_12},  we can write the condition $d(i,j) >0$ as the set of equivalent conditions
\begin{IEEEeqnarray}{lCl}
\label{eq:const1}
f(M,M_i,M_j) -2(M-2r_{\B{\Sigma}}) >0  \IEEEeqnarraynumspace \\
\label{eq:const2}
f(M,M_i,M_j) -2(M_i-r_{\B{\Sigma}}) >0\\
\label{eq:const3}
f(M,M_i,M_j) -2(M_j-r_{\B{\Sigma}}) >0 \\
f(M,M_i,M_j) >0, \label{eq:const4}
\end{IEEEeqnarray}
where $f(M,M_i,M_j)=\max \{  M- r_{\B{\Sigma}}, M_i\} + \max \{  M- r_{\B{\Sigma}}, M_j\}$. 
%\commentFR{say somewhere $M \leq N$}
Then, the upper bound in \eqref{eq:UB_PT_multiclass} is obtained as the solution of the integer optimization problem that aims at minimizing $M=\sum_{i=1}^L M_i$ subject to the constraints \eqref{eq:const1}-\eqref{eq:const4} and $ 0 \leq M_i \leq N- r_{\B{\Sigma}}$.

In the remainder of this appendix, we will show that the solution of such minimization problem is given by $M=\min \{L-1, r_{\B{\Sigma}}+1  \}$, by considering separately two cases. In particular, when $L-1 \leq r_{\B{\Sigma}}$, we can show that the optimal solution is given by $M=\sum_{i=1}^L M_i =L-1$. We first observe that such value represents a feasible solution: in fact, by picking only 1 measurement from $L-1$ out of $L$ null spaces, e.g., by choosing $M_1=\cdots=M_{L-1}=1$ and $M_L=0$, we can immediately prove that all the constraints are verified. Then, we also observe that any solutions for which $M<L-1$ is not feasible: in fact, if $M<L-1$ there exist at least two indexes $k$ and $\ell$ such that $M_k=M_\ell =0$, and therefore at least one of the constraints \eqref{eq:const4} is not verified.

Consider now the case $L-1 > r_{\B{\Sigma}}$. In this case the optimal solution of the minimization problem yields $M=r_{\B{\Sigma}}+1$. In a similar way to the previous case, we start by observing that $M=r_{\B{\Sigma}}+1$ is a feasible solution, which can be achieved by picking 1 measurement from $r_{\B{\Sigma}}+1$ different null space, e.g., by picking $M_1=\cdots = M_{r_{\B{\Sigma}}+1}=1$ and $M_{r_{\B{\Sigma}}+2} =\cdots =M_{L}=0$. Also in this case it is straightforward to prove that all the constraints are verified. Moreover, it is possible to observe that there is not any feasible solution such that $M < r_{\B{\Sigma}}+1$, as $r_{\B{\Sigma}}<L-1$ implies that there exist at least two indexes $k$ and $\ell$ such that $M_k=M_\ell =0$, and, therefore, at least one of the constraints \eqref{eq:const4} is not verified.

\bibliographystyle{IEEEtran}
\bibliography{IEEEabrv,cl_short}

\end{document}